\documentclass[amsmath,amssymb,prl,twocolumn]{revtex4-2}

\usepackage{graphicx,natbib}
\usepackage{dcolumn}
\usepackage{bm}
\usepackage{hyperref}
\usepackage{xcolor}
\usepackage{subfig}

\newcommand{\be}{\begin{equation}}
\newcommand{\ee}{\end{equation}}
\newcommand{\bea}{\begin{eqnarray}}
\newcommand{\eea}{\end{eqnarray}}
\newcommand{\bean}{\begin{eqnarray*}}
\newcommand{\eean}{\end{eqnarray*}}

\newcommand{\prlsection}[1]{\emph{#1.}---}

\begin{document}

\title{Fuzzball Shadows: Emergent Horizons from Microstructure}

\author{Fabio Bacchini}\email{fabio.bacchini@kuleuven.be}
 \affiliation{Centre for mathematical Plasma Astrophysics, Department of Mathematics, Katholieke Universiteit Leuven, Celestijnenlaan 200B, B-3001 Leuven, Belgium}
\author{Daniel R. Mayerson}\email{daniel.mayerson@ipht.fr}\affiliation{Institut de Physique Th\'eorique, Universit\'e Paris-Saclay, CNRS, CEA, Orme des Merisiers 91191, Gif-sur-Yvette CEDEX, France}
\author{Bart Ripperda}\affiliation{Center for Computational Astrophysics, Flatiron Institute, 162 Fifth Avenue, New York, NY 10010, USA}\affiliation{Department of Astrophysical Sciences, Peyton Hall, Princeton University, Princeton, NJ 08544, USA}
\author{Jordy Davelaar}\affiliation{Columbia Astrophysics Laboratory, Columbia University, 550 W 120th St, New York, NY 10027, USA}\affiliation{%
 Center for Computational Astrophysics, Flatiron Institute, 162 Fifth Avenue, New York, NY 10010, USA}\affiliation{Department of Astrophysics/IMAPP, Radboud University Nijmegen, P.O. Box 9010, 6500 GL Nijmegen, The Netherlands}
\author{H\'{e}ctor Olivares}\affiliation{Department of Astrophysics/IMAPP, Radboud University Nijmegen, P.O. Box 9010, 6500 GL Nijmegen, The Netherlands}
\author{Thomas Hertog}\affiliation{Institute for Theoretical Physics, KU Leuven, Celestijnenlaan 200D, B-3001 Leuven, Belgium}
\author{Bert Vercnocke}\affiliation{Institute for Theoretical Physics, KU Leuven, Celestijnenlaan 200D, B-3001 Leuven, Belgium}

\begin{abstract}
We study the physical properties of four-dimensional, string-theoretical, horizonless ``fuzzball'' geometries by imaging their shadows. Their microstructure traps light rays straying near the would-be horizon on long-lived, highly redshifted chaotic orbits. In fuzzballs sufficiently near the scaling limit this creates a shadow much like that of a black hole, while avoiding the paradoxes associated with an event horizon. Observations of the shadow size and residual glow can potentially discriminate between fuzzballs away from the scaling limit and alternative models of black compact objects.
\end{abstract}

\maketitle

\prlsection{{\bf Introduction}}
The recent images of the shadow of the supermassive black hole (BH) at the center of M87 taken by the Event Horizon Telescope (EHT) \cite{EHT2019a} open up the exciting prospect to explore the strong-gravity environment near BHs with horizon-scale resolution. 

It has long been believed that the vicinity of the horizon is well described by BH solutions of classical general relativity. However, many theorists who study the quantum evolution of BHs have now concluded that there must be modifications to BHs extending to horizon scales (see e.g. \cite{Mathur:2009hf}), in order to resolve the information paradox that stems from the discovery of Hawking radiation and BH evaporation. A variety of proposals have been made, among which exotic compact objects (ECOs) \cite{Cardoso_2017njb,Cardoso:2017cqb,Cardoso:2019rvt} such as boson stars \cite{Kaup:1968zz} or gravastars \cite{Mazur:2004fk}, firewalls \cite{Almheiri:2012rt}, new non-local interactions \cite{Giddings:2006sj}, and fuzzball geometries \cite{Mathur:2005zp}. Since shadows of compact black objects are sensitive to the object's properties in the near-horizon region (see e.g. \cite{Vincent:2015xta,Grould:2017rzz,Olivares:2018abq,Giddings:2016btb}), their observation offers a promising route to differentiate between some of these proposed BH alternatives. 
Here we concentrate on the subset of fuzzballs known as ``microstate geometries''  \cite{Bena:2013dka,Bena:2015dpt,Bena:2017xbt,Heidmann:2019xrd}. These are horizonless solutions of low-energy string theory that describe compact objects with an intricate geometric and topological inner structure and which can be thought of as coherent ``microstates'' corresponding to BHs. Therefore, in contrast to the ECO models usually employed in phenomenological studies, fuzzballs admit a firm embedding in string theory; as such, they are guaranteed to have a UV-complete description and they will not lie within the inadmissible string-theory swampland \cite{Ooguri:2006in,Li:2021gbg}. Their phenomenology has recently begun to be explored; see e.g.\ \cite{Hertog:2017vod,Mayerson:2020tpn} and also \cite{Bena:2020see,Bena:2020uup,Bianchi:2020bxa,Bianchi:2020miz}. 

We study the behavior of null geodesics (light rays) in four-dimensional fuzzball geometries 
and use this to obtain the first images of fuzzball shadows as they would be perceived by a distant observer.
To this end we first construct a novel axisymmetric scaling solution, the ``ring fuzzball''. Our analysis of this confirms that, as far as the behavior of geodesics is concerned, fuzzballs in the scaling limit can resemble BH geometries arbitrarily well. It also reveals the key features and mechanisms of fuzzballs through which such phenomenological horizon behavior emerges.

Even though the geometries we study here are supersymmetric and have various limitations \cite{Mayerson:2020tpn}, we expect that the overall properties of their shadows presented here qualitatively capture those of shadows of more general (as yet unconstructed) fuzzballs which correspond to realistic BHs.

\prlsection{{\bf Geometry}}
We consider a class of four-dimensional multi-center solutions in supergravity \cite{Bates:2003vx,Denef:2000nb,Berglund:2005vb} with metric
 \be \label{eq:ds2multicenter} ds^2 = - \mathcal{Q}^{-1/2}(dt +  \omega)^2 + \mathcal{Q}^{1/2}\left( dx^2+dy^2+dz^2 \right),\ee
 which are smooth and horizonless when uplifted to five dimensions (and have controlled singularities in four dimensions). These solutions are completely determined by harmonic functions $(V,K,L,M)$ on the flat $\mathbb{R}^3$ basis spanned by $(x,y,z)$ \cite{Gauntlett:2004qy, Bena:2005ni}.
 The metric warp factor is then given by
\begin{align}
\label{eq:quarticinvdef} \mathcal{Q} & = Z^3 V - \mu^2 V^2, ~ Z = L +  \frac{K^2}{V}, ~
 \mu = M + \frac{3KL}{2V} + \frac{K^3}{V^2}.
\end{align}
Finally, the rotation one-form $\omega$ is determined by
\be \label{eq:omegadiff} *_3 d\omega = V dM -M dV + \frac32\left( K dL - L dK\right),\ee
where $*_3$ is the Hodge star with respect to the flat $\mathbb{R}^3$ basis.

In this Letter, we consider microstate geometries inspired by the ``black-hole deconstruction'' paradigm \cite{Denef:2007yt,Levi:2009az,Raeymaekers:2015sba}, which consist of a pair of fluxed D6-$\overline{\text{D6}}$ brane centers surrounded by a halo of D0 brane centers
\cite{Denef:2007yt,Maldacena:1997de,Gaiotto:2004ij}. The harmonic functions are then given by
\begin{align} \label{eq:harmfuncring}
\nonumber V& = \left(\frac{1}{r_1} - \frac{1}{r_2}\right), ~  M = -\frac12+\frac{P^3}{2}\left(\frac{1}{r_1} + \frac{1}{r_2}\right) + M_{D0},\\
K &= 1+P\left(\frac{1}{r_1} + \frac{1}{r_2}\right), ~
L = -P^2\left(\frac{1}{r_1} - \frac{1}{r_2}\right).
\end{align}
The ``centers'' at $\vec{r}_{1}$ and $\vec{r}_2$ are on the $z$-axis at $z=\pm l/2$ and are fluxed D6 branes in a ten-dimensional perspective.
For $M_{D0}$, with any number of centers with total charge $-q_0$ (with $q_0>0$) distributed on the circle $x^2+y^2=R^2$ at $z=0$, the so-called bubble equations (i.e.\ regularity conditions on the metric) \cite{Bates:2003vx,Bena:2007kg} relate $l$ and $R$ as
\begin{align}
    l &= 8P^3 \lambda, & R &= 2\lambda\sqrt{\frac{q_0^2}{(1-(1-3P^2)\lambda)^2}-4P^6} .
\end{align}  
This is a one-parameter family of solutions entirely determined by $\lambda>0$. We will show that $\lambda$ can be effectively tuned to obtain a correspondence between microstate geometries and classical BHs.

Instead of considering point singularities, we employ a ``smeared'' configuration of uniform charge density on the entire circle of radius $R$, giving
\be\label{eq:MD0ring} M_{D0} = \frac{-2q_0}{\pi\sqrt{r^2 + R^2 +2r R \sin\theta}} \mathcal{K}\left(\frac{4r R \sin\theta}{r^2+R^2+2r R \sin\theta}\right),\ee
where $\mathcal{K}(\cdot)$ is the complete elliptic integral of the first kind and we used spherical coordinates $(r,\theta,\phi)$ on the flat $\mathbb{R}^3$ base. In this case, $\omega=\omega_\phi d\phi$ and (\ref{eq:ds2multicenter}) is axisymmetric and equatorially symmetric.
This microstate geometry, determined by \eqref{eq:harmfuncring} and \eqref{eq:MD0ring}, is the ``ring fuzzball'' which we will study using ray tracing.

The ring fuzzball solution allows for the ``scaling limit'' $\lambda\rightarrow 0$, in which it approaches the metric of the static, supersymmetric BH given by \eqref{eq:ds2multicenter} with
\begin{align}
 \label{eq:BHharmfunc}   V &=L= 0, & K &=1+ \frac{2P}{r},  & M =-\frac12+ \frac{P^3 - q_0}{r}.
\end{align}
This BH has mass $2M_{BH}=3P+q_0-P^3$ and a horizon at $r=0$ with area $A_{BH} = 16\pi \sqrt{P^3(q_0-P^3)}$. The angular momentum of the BH vanishes while the ring fuzzball has  $J=2P^3(1-3P^2)\lambda$. (Our units are such that $G_4=1/16$ \cite{Bena:2007kg}.)
Note that this ring fuzzball represents the first ever explicitly constructed (exactly) scaling solution that is also axisymmetric.

\prlsection{{\bf Ray Tracing}}
We explore the ring fuzzball geometry by tracing geodesics with the ray-tracing code \textsc{Raptor} \cite{bronzwaer2018,davelaar2018b,bronzwaer2020}. Ray tracing allows for visualizing the environment of compact objects (e.g.\ \cite{bacchini2018a}). The observer is represented by a camera at a specified position with $10^6$ pixels.
Geodesics are numerically integrated backwards in time from the camera to where they originated.
The integration is halted when a geodesic reaches an outer ``celestial sphere'', with chosen radius equal to the camera's radial location. To highlight lensing patterns, each geodesic is assigned a color according to the quadrant (each of equal angular width in $\phi$) of the sphere that it originates from \cite{Bohn:2014xxa}.
If a geodesic does not reach the celestial sphere within a predetermined maximum integration time (corresponding to an observation window for a physical observer), we assign the color black to it; this includes geodesics falling in to horizons.

In our four-dimensional model, the D6-$\overline{\text{D6}}$ centers and the D0 ring are (naked) singularities; however, in our calculations no geodesics ever approach a singularity close enough (i.e.\ $<10^{-14}$ in $\mathbb{R}^3$ coordinate distance) to require the application of boundary conditions at the singularities.
 (Nevertheless, such boundary conditions can be derived from higher-dimensional string-theoretical embedding \cite{ourfollowup}.)

For each geodesic, we track the position on the celestial sphere, the total elapsed (coordinate) time on its trajectory, as well as:
\begin{itemize}
    \item \emph{Redshift}, measured as the minimum value of $\sqrt{-g_{tt}}=\mathcal{Q}^{-1/4}$ that a geodesic encounters on its path. This provides a measure of how much energy a geodesic would lose in escaping from the gravity well.
    
    \item \emph{Curvature}, measured as the maximum value of  $K=R^{\mu\nu\rho\sigma}R_{\mu\nu\rho\sigma}$ that the geodesic encounters on its path. This provides an estimate of tidal stress a geodesic encounters.
\end{itemize}
We have also tracked the actual tidal stresses along geodesics (in the spirit of higher-dimensional studies of tidal forces in microstate geometries \cite{Tyukov:2017uig,Bena:2018mpb,Bena:2020iyw,Martinec:2020cml}). However, we found that these quantities provide the same information given by $K$; we will discuss these further in \cite{ourfollowup}.

\begin{figure*}
\centering
\includegraphics[width=1\textwidth,trim=0 0 8cm 0, clip]{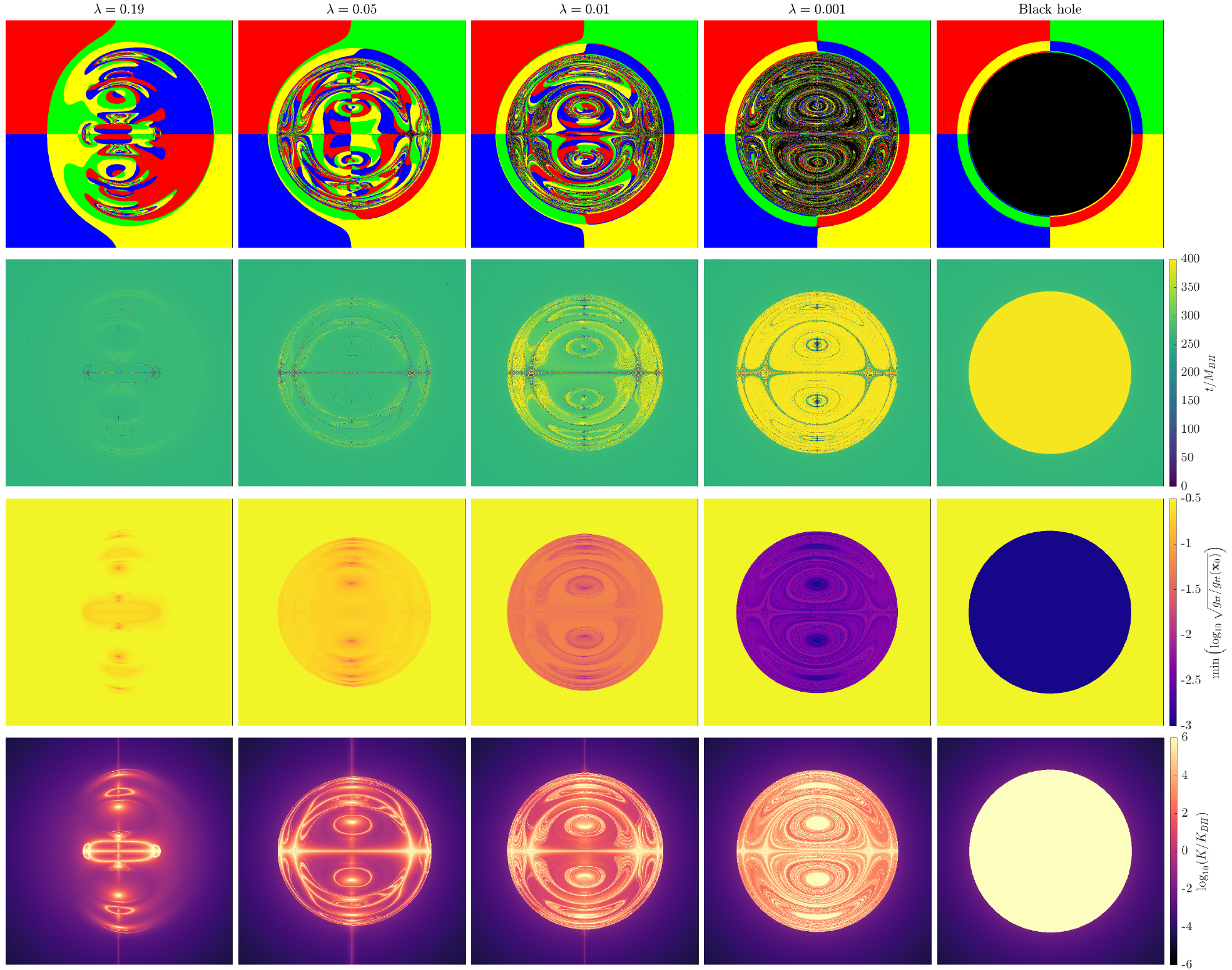}
\caption{From left to right columns: visualizations of diagnostics for the ring fuzzball ($P=2$ and $q_0=50$) for decreasing values of $\lambda$, compared to the $\lambda\rightarrow0$ BH in the rightmost column. Rows from top to bottom: four-color screen indicating the portion of the celestial sphere from which geodesics originate; coordinate time $t$ elapsed at the end of the numerical integration (normalized to $M_{BH}$); strongest redshift experienced by the geodesic (normalized to the redshift at the camera position); strongest curvature encountered (normalized to $K$ at the BH horizon).}
\label{fig:ST3Cr_lambda_comparison}
\end{figure*}

\prlsection{{\bf Results}}
Our results are presented in Figure \ref{fig:ST3Cr_lambda_comparison} for the ring fuzzball defined by \eqref{eq:harmfuncring} and \eqref{eq:MD0ring} with the representative choice $P=2$ and $q_0=50$, and for values $\lambda=0.19, 0.05, 0.01, 0.001$. (The bubble equations require $\lambda\leq 17/88\approx 0.193$.)
For comparison, we also show results for the BH defined by \eqref{eq:BHharmfunc} (also with $P=2,q_0=50$), which corresponds to the $\lambda\rightarrow 0$ limit of the ring fuzzball. From top to bottom row, we show the light bending (via the four-color screen visualization), the total elapsed coordinate time, the redshift, and the curvature. In all cases the camera is located at $\textbf{x}_0=(250,0,0)$, pointed towards the origin, and oriented such that the $z$-axis is vertical. (The camera is then at approximately the same physical distance to the object in all images.)

As $\lambda$ decreases to 0, the microstate geometry should resemble the BH increasingly well \cite{Bena:2006kb,Bena:2007kg,Bena:2007qc}. Here we observe the explicit mechanism by which the ring fuzzball, despite possessing no horizon, does so. First, as $\lambda$ decreases, the four-color screen pictures reveal that increasingly many geodesics traveling near the would-be horizon scale follow chaotic paths. They are scattered across the celestial sphere and form fewer coherent geometric structures (as indicated by the larger chaotically colored regions). Second, the elapsed-time pictures confirm that these complicated chaotic orbits are long-lived; moreover, as $\lambda$ decreases, an increasing amount of geodesics do not escape the near-center region within the observation window (as indicated by the presence of more black pixels in the four-color screen pictures).

\begin{figure*}
\centering
\includegraphics[width=1\textwidth,trim=0 6.75cm 0.2cm 0, clip]{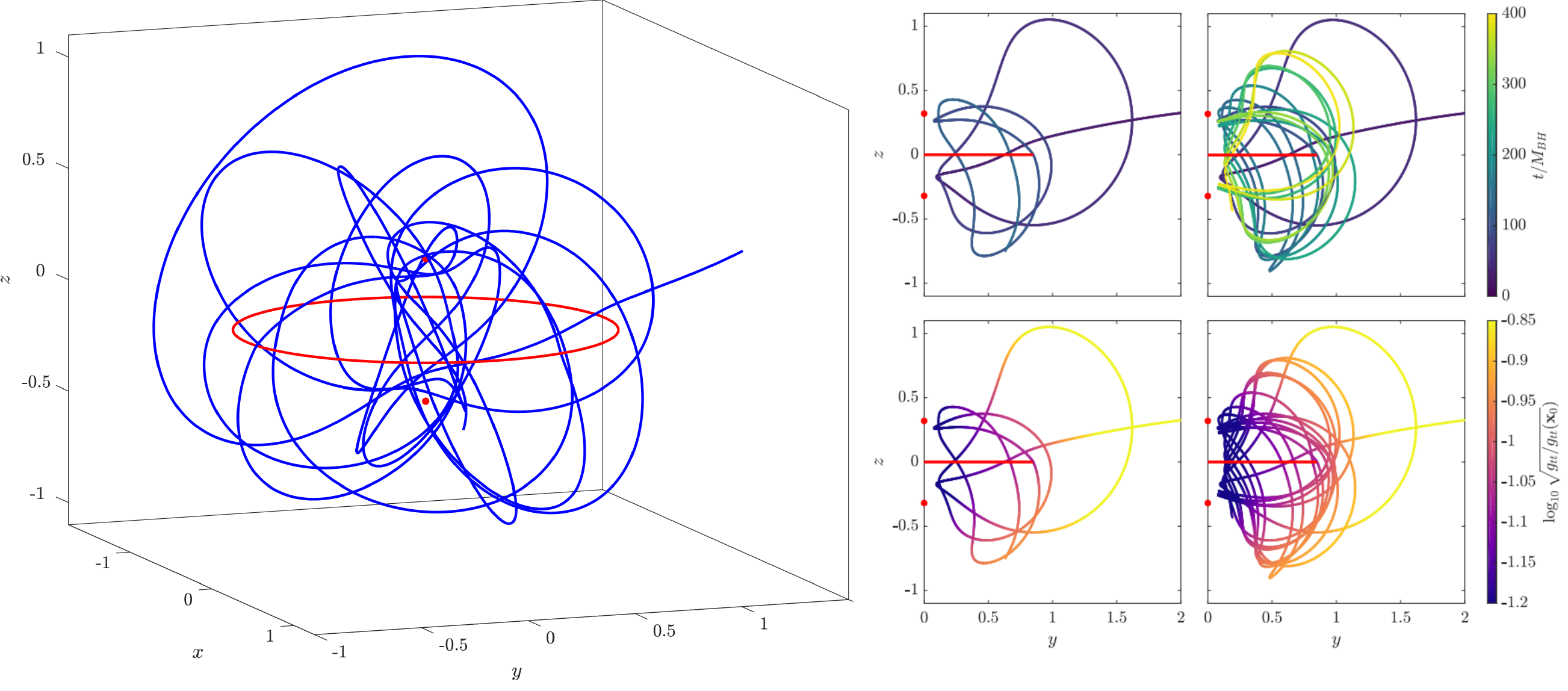}
\caption{Left panel: three-dimensional visualization of a representative geodesic in the ring fuzzball geometry with $P=2,q_0=50$ and $\lambda=0.01$. The on-axis centers and the charged ring are shown in red. Right panels: a short portion (left column) and the full trajectory (right column) projected on the $\phi=\pi/2$ plane, colored by elapsed time (top) and local redshift (bottom).}
\label{fig:geodesictrajectories}
\end{figure*}

Finally, the redshift and curvature plots show that geodesics in such long-lived, almost-trapped chaotic orbits also experience extremely strong redshift and encounter large spacetime curvature. When including backreaction, photons on such trajectories will interact with and lose energy to the fuzzball microstructure, leading to effective scrambling behavior as expected in a BH. Thus, such photons -- even if they escape after their long, chaotic orbits -- will be significantly redshifted and fall outside any detectable wavelength range. It is therefore the combined effect of long-term trapping and strong redshift that makes fuzzball microstates exhibit an emerging horizon behavior and ultimately mimic a classical BH.

We also note that the geodesics in the ring fuzzball metric encode information on the inner structure of the BH it mimics. The curvature plots in the last row of Figure \ref{fig:ST3Cr_lambda_comparison} indicate that one can interpret the microstate geometry as ``resolving'' the curvature singularity of a BH into constituent pieces. In a higher-dimensional, string-theoretical embedding, these pieces would even be completely smooth without curvature divergences. 

The geodesic trapping mechanism is further shown explicitly in Figure \ref{fig:geodesictrajectories}, where we plot a representative geodesic for the $\lambda=0.01$ case (also for $P=2,q_0=50$). This shows that the geodesic is trapped in the near-center region and experiences strong redshift as it bounces between the two on-axis centers and the charge ring. Further, the chaotic and long-lived nature of these orbits is apparent; geodesics may remain trapped far longer than the observation window.

For higher-dimensional microstate geometries, ``chaotic trapping'' behavior of geodesics was anticipated in \cite{Bianchi:2017sds,Bianchi:2018kzy}, and was shown to be related to the presence of so-called ``evanescent ergosurfaces'' (timelike surfaces of infinite redshift); these surfaces are also linked to instabilities in these geometries \cite{Eperon:2016cdd,Keir:2016azt,Marolf:2016nwu,Eperon:2017bwq}. There are no evanescent ergosurfaces in four-dimensional microstate geometries; our results are thus the first explicit indication and exploration of such trapping effects in this context. Note that a chaotic trapping mechanism was observed in lensing in boson star ECO models \cite{Cunha:2015yba,Cunha:2016bjh}; however, we note that, in order to form an emergent horizon, null geodesics must also explore regions of extreme curvature and redshift. The results presented here show that all of these features appear naturally in a model based on microstructure, whereas it is unlikely that they can arise in stable boson stars.

\begin{figure}
\centering
\includegraphics[width=1\columnwidth,trim=0 0 12.25cm 0, clip]{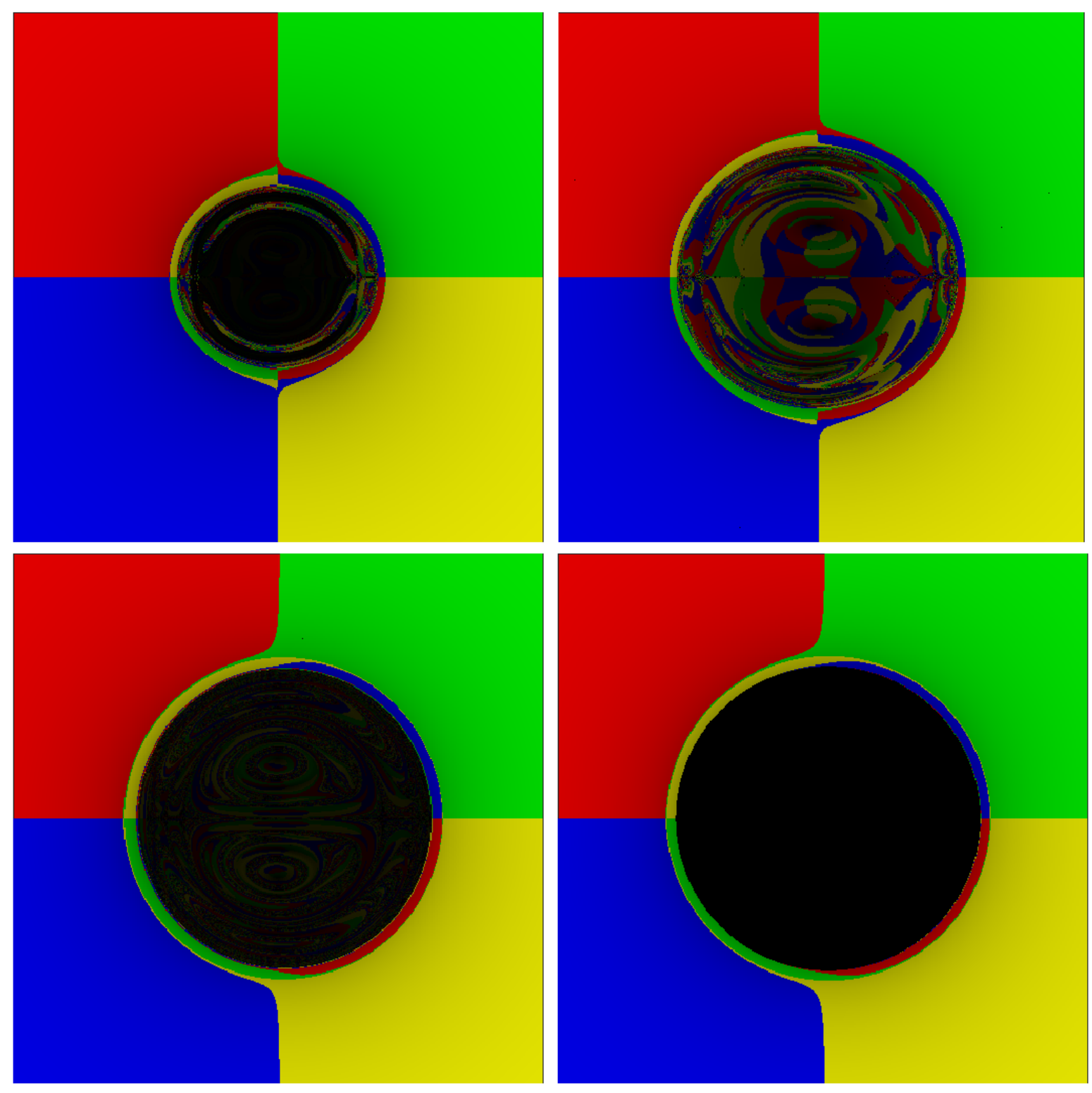}
\caption{Visualization (see text) of ring fuzzballs with $P=1/4,\lambda=0.1$ (top left), $P=307/500,\lambda=0.3$ (top right), $P=7/8,\lambda=0.11$ (bottom left), compared to an extremal Kerr-Newman BH (bottom right) of equal mass (and same angular momentum as the bottom-left fuzzball). The two bottom pictures exhibit minimal but non-zero differences.}
\label{fig:Kerrcomparison}
\end{figure}

To further make the phenomenological possibilities of these microstate geometries explicit, in Figure \ref{fig:Kerrcomparison} we compare the effective shadows of three ring fuzzballs with that of an extremal Kerr-Newman (KN) BH, all with mass $M=7/4$. (Note that the charge of this BH can be interpreted as ``dark charge'' that only interacts gravitationally with standard-model particles and is not ruled out by current observations \cite{Cardoso:2016olt,Bozzola:2020mjx,ourfollowupKN}.
Note that we are focusing on the comparison of microstate geometry shadows with those of BHs, and thus making the simplifying assumption that extremal KN with dark charge is not ruled out for a given BH mass and angular momentum \cite{ourfollowupKN}. A more detailed, realistic comparison would be to allow the mass and charge of the (KN) BH to vary within the allowed range of EHT shadow observations of M87* and especially to compare explicitly with (uncharged) Kerr.) We construct the four-color pictures, and darken each pixel according to the (maximal) redshift the geodesic has encountered on its path. We assume that light will scatter off the microstructure (due to e.g.\ the large tidal stresses encountered) before they can escape from the near-center region, making the redshift a reasonable proxy for the energy loss of the photon and therefore the darkness of the resulting image.
In this way, the observed image of the microstate geometry can appear as dark as that of a BH. Figure \ref{fig:Kerrcomparison} clearly shows that, depending on the value of $P,\lambda$, fuzzballs can mimic BHs to different degrees. In particular, fuzzball shadows can appear very dark, but with a much smaller area (top left); or, while the shadow size can be within observational uncertainty bounds \cite{EHT2019a}, a weak redshift may make fuzzballs appear too bright (top right). With the appropriate parameters, fuzzballs can also appear almost indistinguishable from actual BHs (bottom left).
Quantitatively, if we determine $q_0$ by keeping the mass fixed as we vary $P$, then the values $0.695\lesssim P\lesssim 0.912$ correspond to solutions with an effective horizon area within 10\% of that of the BH; this is approximately the uncertainty in the shadow size of M87* measured by the EHT \cite{EHT2019a}. This bound rules out the top-left panel in Figure \ref{fig:Kerrcomparison}. Furthermore, for M87* the measured upper bound on the brightness at the center of the shadow is $<10\%$ of the average image brightness \cite{EHT2019a}. If we assume this implies a redshift of $\sqrt{-g_{tt}}<0.1$ at the center of the fuzzball shadow, the top-right panel in Figure \ref{fig:Kerrcomparison} is just within this acceptable bound, and the bottom-left panel is well within this bound.

\prlsection{{\bf Discussion}}
We have studied the physical properties of fuzzballs, both as interesting geometries in their own right and as phenomenological alternatives to BHs, by numerically calculating and imaging the behavior of null geodesics in a class of microstate geometries. The images presented here are the first visualizations of fuzzball geometries. They confirm several physical properties of microstate geometries which had generally been anticipated but never verified, such as their trapping behavior. As the ring fuzzball approaches the BH limit, its microstructure induces increasingly chaotic motion of geodesics straying in the near-center region. 
Infalling geodesics that reach the microstructure will be heavily blueshifted and subsequently backreact with the structure and/or be heavily scattered. The light that will emanate from this region will then be too
redshifted to be detectable.
In this way, the microstructure conspires to create a shadow very much like that of a BH, while avoiding the paradoxes associated with an event horizon. Note that the physical mechanisms we report for geodesics in the ring fuzzball are entirely generic; we have studied other scaling microstate geometries, for example non-axisymmetric ones, with entirely analogous results \cite{ourfollowup}. However, note that other ECOs may not share these mechanisms. For example, Solodukhin-type wormholes \cite{Bueno:2017hyj} can be arbitrarily compact, but do not have regions of large tidal stress; they will also appear dark, but this is due to the escaping of light rays into a second asymptotic region.

Our results indicate that fuzzballs sufficiently near the scaling limit yield a theoretically appealing and phenomenologically viable BH alternative. Vice versa, observations of a faint residual glow in the object's shadow, or of its shadow's size compared to its mass, have the potential to discriminate between BHs and fuzzballs somewhat away from the scaling limit. These should therefore be prime targets for current and future imaging missions such as the EHT. 

This motivates a more detailed investigation of the characteristics that differentiate microstate geometries from BHs through the intricate structure of their shadows. This will require not only further advances in the construction of more realistic fuzzballs but also an accurate modeling of the plasma in a realistic accretion disk, coupled to full radiative transfer methods. This approach has been employed for BH geometries considered by the EHT \cite{EHT2019e}, as well as for more exotic objects \cite{mizuno2018,Olivares:2018abq}, and we intend to report on this elsewhere \cite{ourfollowup,ourfollowupKN}. Another generalization of our analysis could be to include explicit source terms in the geodesic equation, in order to model the energy exchange between light rays and the geometry; this would further quantify  the emergence of horizon-like shadows. These source terms can be modeled after stringy tidal excitations \cite{Martinec:2020cml}. In a further step, the subsequent evolution of the microstate geometry can be studied, using e.g.\ inspiration from the quantum tunneling dynamics studied in \cite{Bena:2015dpt} to model this evolution as a dynamical tendency to form additional centers in the microstate geometry (which itself could be an effect similar to the instability of large extremal black holes to split into smaller ones, as expected within the weak gravity conjecture \cite{Arkani-Hamed:2006emk,Loges:2019jzs}). It would also especially be interesting to understand the relation with the (classical) microstate geometry (in)stabilities of \cite{Eperon:2016cdd,Keir:2016azt,Marolf:2016nwu,Eperon:2017bwq} and the additional role these may play in the interactions and subsequent evolution and dynamics of the geometry.

Moreover, the chaotic behaviour of geodesics on the would-be horizon scales also suggests that gravitational waves propagating in this region will similarly be chaotically dispersed. Therefore one expects the resulting gravitational wave signal that leaks out to be chaotic and dispersed over extended time intervals. This suggests that the post-ringdown phase of gravitational waves generated in fuzzball mergers will not exhibit a markedly clear echo structure (see \cite{Dimitrov:2020txx,Mayerson:2020tpn}).\\

\begin{acknowledgements}
We would like to thank Iosif Bena, Vitor Cardoso, Nejc Ceplak, Bogdan Ganchev, Shaun Hampton, Carlos Herdeiro, Tom Lemmens, Yixuan Li, Marina Martinez, Ben Niehoff, Nicholas Warner for useful discussions, Tom Lemmens for collaboration and suggestions from closely related work, and Iosif Bena for comments and suggestions on a preliminary draft of this Letter.
The computational resources and services used in this work were provided by the VSC (Flemish Supercomputer Center), funded by the Research Foundation -- Flanders (FWO) and the Flemish Government -- department EWI.
FB is supported by a Junior PostDoctoral Fellowship (grant number 12ZW220N) from Research Foundation -- Flanders (FWO).
DRM is supported by ERC Advanced Grant 787320 - QBH Structure and ERC Starting Grant 679278 - Emergent-BH.
BR is supported by a Joint Princeton/Flatiron Postdoctoral Fellowship. Research at the Flatiron Institute is supported by the Simons Foundation.
JD is supported by NASA grant NNX17AL82G.
HO is supported by a Virtual Institute of Accretion (VIA) postdoctoral fellowship from the Netherlands Research School for Astronomy (NOVA).
TH and BV were supported by ERC-CoG HoloQosmos 616732, C16/16/005 KU Leuven research grant, the FWO research projects GCD-C3714-G.0011.12 (Odysseus) and G092617N, and the COST action CA16104 GWVerse CA16104.

FB and DRM contributed equally to this work.

\end{acknowledgements}


\begin{thebibliography}{66}%
\makeatletter
\providecommand \@ifxundefined [1]{%
 \@ifx{#1\undefined}
}%
\providecommand \@ifnum [1]{%
 \ifnum #1\expandafter \@firstoftwo
 \else \expandafter \@secondoftwo
 \fi
}%
\providecommand \@ifx [1]{%
 \ifx #1\expandafter \@firstoftwo
 \else \expandafter \@secondoftwo
 \fi
}%
\providecommand \natexlab [1]{#1}%
\providecommand \enquote  [1]{``#1''}%
\providecommand \bibnamefont  [1]{#1}%
\providecommand \bibfnamefont [1]{#1}%
\providecommand \citenamefont [1]{#1}%
\providecommand \href@noop [0]{\@secondoftwo}%
\providecommand \href [0]{\begingroup \@sanitize@url \@href}%
\providecommand \@href[1]{\@@startlink{#1}\@@href}%
\providecommand \@@href[1]{\endgroup#1\@@endlink}%
\providecommand \@sanitize@url [0]{\catcode `\\12\catcode `\$12\catcode
  `\&12\catcode `\#12\catcode `\^12\catcode `\_12\catcode `\%12\relax}%
\providecommand \@@startlink[1]{}%
\providecommand \@@endlink[0]{}%
\providecommand \url  [0]{\begingroup\@sanitize@url \@url }%
\providecommand \@url [1]{\endgroup\@href {#1}{\urlprefix }}%
\providecommand \urlprefix  [0]{URL }%
\providecommand \Eprint [0]{\href }%
\providecommand \doibase [0]{https://doi.org/}%
\providecommand \selectlanguage [0]{\@gobble}%
\providecommand \bibinfo  [0]{\@secondoftwo}%
\providecommand \bibfield  [0]{\@secondoftwo}%
\providecommand \translation [1]{[#1]}%
\providecommand \BibitemOpen [0]{}%
\providecommand \bibitemStop [0]{}%
\providecommand \bibitemNoStop [0]{.\EOS\space}%
\providecommand \EOS [0]{\spacefactor3000\relax}%
\providecommand \BibitemShut  [1]{\csname bibitem#1\endcsname}%
\let\auto@bib@innerbib\@empty
\bibitem [{\citenamefont {{EHT~Collaboration}}(2019{\natexlab{a}})}]{EHT2019a}%
  \BibitemOpen
  \bibfield  {author} {\bibinfo {author} {\bibnamefont {{EHT~Collaboration}}},\
  }\bibfield  {title} {\bibinfo {title} {{First M87 Event Horizon Telescope
  Results. I. The Shadow of the Supermassive Black Hole}},\ }\href
  {https://doi.org/10.3847/2041-8213/ab0ec7} {\bibfield  {journal} {\bibinfo
  {journal} {ApJL}\ }\textbf {\bibinfo {volume} {875}},\ \bibinfo {eid} {L1}
  (\bibinfo {year} {2019}{\natexlab{a}})}\BibitemShut {NoStop}%
\bibitem [{\citenamefont {Mathur}(2009)}]{Mathur:2009hf}%
  \BibitemOpen
  \bibfield  {author} {\bibinfo {author} {\bibfnamefont {S.~D.}\ \bibnamefont
  {Mathur}},\ }\bibfield  {title} {\bibinfo {title} {{The Information paradox:
  A Pedagogical introduction}},\ }\href@noop {} {\bibfield  {journal} {\bibinfo
   {journal} {Class. Quant. Grav.}\ }\textbf {\bibinfo {volume} {26}},\
  \bibinfo {pages} {224001} (\bibinfo {year} {2009})}\BibitemShut {NoStop}%
\bibitem [{\citenamefont {Cardoso}\ and\ \citenamefont
  {Pani}(2017{\natexlab{a}})}]{Cardoso_2017njb}%
  \BibitemOpen
  \bibfield  {author} {\bibinfo {author} {\bibfnamefont {V.}~\bibnamefont
  {Cardoso}}\ and\ \bibinfo {author} {\bibfnamefont {P.}~\bibnamefont {Pani}},\
  }\bibfield  {title} {\bibinfo {title} {{The observational evidence for
  horizons: from echoes to precision gravitational-wave physics}},\ }\href@noop
  {} {\bibfield  {journal} {\bibinfo  {journal} {arXiv:1707.030211}\ }
  (\bibinfo {year} {2017}{\natexlab{a}})}\BibitemShut {NoStop}%
\bibitem [{\citenamefont {Cardoso}\ and\ \citenamefont
  {Pani}(2017{\natexlab{b}})}]{Cardoso:2017cqb}%
  \BibitemOpen
  \bibfield  {author} {\bibinfo {author} {\bibfnamefont {V.}~\bibnamefont
  {Cardoso}}\ and\ \bibinfo {author} {\bibfnamefont {P.}~\bibnamefont {Pani}},\
  }\bibfield  {title} {\bibinfo {title} {{Tests for the existence of black
  holes through gravitational wave echoes}},\ }\href@noop {} {\bibfield
  {journal} {\bibinfo  {journal} {Nature Astron.}\ }\textbf {\bibinfo {volume}
  {1}},\ \bibinfo {pages} {586} (\bibinfo {year}
  {2017}{\natexlab{b}})}\BibitemShut {NoStop}%
\bibitem [{\citenamefont {Cardoso}\ and\ \citenamefont
  {Pani}(2019)}]{Cardoso:2019rvt}%
  \BibitemOpen
  \bibfield  {author} {\bibinfo {author} {\bibfnamefont {V.}~\bibnamefont
  {Cardoso}}\ and\ \bibinfo {author} {\bibfnamefont {P.}~\bibnamefont {Pani}},\
  }\bibfield  {title} {\bibinfo {title} {{Testing the nature of dark compact
  objects: a status report}},\ }\href@noop {} {\bibfield  {journal} {\bibinfo
  {journal} {Living Rev. Rel.}\ }\textbf {\bibinfo {volume} {22}},\ \bibinfo
  {pages} {4} (\bibinfo {year} {2019})}\BibitemShut {NoStop}%
\bibitem [{\citenamefont {Kaup}(1968)}]{Kaup:1968zz}%
  \BibitemOpen
  \bibfield  {author} {\bibinfo {author} {\bibfnamefont {D.~J.}\ \bibnamefont
  {Kaup}},\ }\bibfield  {title} {\bibinfo {title} {{Klein-Gordon Geon}},\
  }\href {https://doi.org/10.1103/PhysRev.172.1331} {\bibfield  {journal}
  {\bibinfo  {journal} {Phys. Rev.}\ }\textbf {\bibinfo {volume} {172}},\
  \bibinfo {pages} {1331} (\bibinfo {year} {1968})}\BibitemShut {NoStop}%
\bibitem [{\citenamefont {Mazur}\ and\ \citenamefont
  {Mottola}(2004)}]{Mazur:2004fk}%
  \BibitemOpen
  \bibfield  {author} {\bibinfo {author} {\bibfnamefont {P.~O.}\ \bibnamefont
  {Mazur}}\ and\ \bibinfo {author} {\bibfnamefont {E.}~\bibnamefont
  {Mottola}},\ }\bibfield  {title} {\bibinfo {title} {{Gravitational vacuum
  condensate stars}},\ }\href {https://doi.org/10.1073/pnas.0402717101}
  {\bibfield  {journal} {\bibinfo  {journal} {Proc. Nat. Acad. Sci.}\ }\textbf
  {\bibinfo {volume} {101}},\ \bibinfo {pages} {9545} (\bibinfo {year}
  {2004})},\ \Eprint {https://arxiv.org/abs/gr-qc/0407075}
  {arXiv:gr-qc/0407075} \BibitemShut {NoStop}%
\bibitem [{\citenamefont {Almheiri}\ \emph {et~al.}(2013)\citenamefont
  {Almheiri}, \citenamefont {Marolf}, \citenamefont {Polchinski},\ and\
  \citenamefont {Sully}}]{Almheiri:2012rt}%
  \BibitemOpen
  \bibfield  {author} {\bibinfo {author} {\bibfnamefont {A.}~\bibnamefont
  {Almheiri}}, \bibinfo {author} {\bibfnamefont {D.}~\bibnamefont {Marolf}},
  \bibinfo {author} {\bibfnamefont {J.}~\bibnamefont {Polchinski}},\ and\
  \bibinfo {author} {\bibfnamefont {J.}~\bibnamefont {Sully}},\ }\bibfield
  {title} {\bibinfo {title} {{Black Holes: Complementarity or Firewalls?}},\
  }\href@noop {} {\bibfield  {journal} {\bibinfo  {journal} {JHEP}\ }\textbf
  {\bibinfo {volume} {02}}\bibinfo  {number} { (2013)},\ \bibinfo {pages}
  {062}}\BibitemShut {NoStop}%
\bibitem [{\citenamefont {Giddings}(2006)}]{Giddings:2006sj}%
  \BibitemOpen
\bibfield  {number} {  }\bibfield  {author} {\bibinfo {author} {\bibfnamefont
  {S.~B.}\ \bibnamefont {Giddings}},\ }\bibfield  {title} {\bibinfo {title}
  {{Black hole information, unitarity, and nonlocality}},\ }\href
  {https://doi.org/10.1103/PhysRevD.74.106005} {\bibfield  {journal} {\bibinfo
  {journal} {Phys. Rev. D}\ }\textbf {\bibinfo {volume} {74}},\ \bibinfo
  {pages} {106005} (\bibinfo {year} {2006})},\ \Eprint
  {https://arxiv.org/abs/hep-th/0605196} {arXiv:hep-th/0605196} \BibitemShut
  {NoStop}%
\bibitem [{\citenamefont {Mathur}(2005)}]{Mathur:2005zp}%
  \BibitemOpen
  \bibfield  {author} {\bibinfo {author} {\bibfnamefont {S.~D.}\ \bibnamefont
  {Mathur}},\ }\bibfield  {title} {\bibinfo {title} {{The Fuzzball proposal for
  black holes: An Elementary review}},\ }\href
  {https://doi.org/10.1002/prop.200410203} {\bibfield  {journal} {\bibinfo
  {journal} {Fortsch. Phys.}\ }\textbf {\bibinfo {volume} {53}},\ \bibinfo
  {pages} {793} (\bibinfo {year} {2005})},\ \Eprint
  {https://arxiv.org/abs/hep-th/0502050} {arXiv:hep-th/0502050} \BibitemShut
  {NoStop}%
\bibitem [{\citenamefont {Vincent}\ \emph {et~al.}(2016)\citenamefont
  {Vincent}, \citenamefont {Meliani}, \citenamefont {Grandclement},
  \citenamefont {Gourgoulhon},\ and\ \citenamefont {Straub}}]{Vincent:2015xta}%
  \BibitemOpen
  \bibfield  {author} {\bibinfo {author} {\bibfnamefont {F.~H.}\ \bibnamefont
  {Vincent}}, \bibinfo {author} {\bibfnamefont {Z.}~\bibnamefont {Meliani}},
  \bibinfo {author} {\bibfnamefont {P.}~\bibnamefont {Grandclement}}, \bibinfo
  {author} {\bibfnamefont {E.}~\bibnamefont {Gourgoulhon}},\ and\ \bibinfo
  {author} {\bibfnamefont {O.}~\bibnamefont {Straub}},\ }\bibfield  {title}
  {\bibinfo {title} {{Imaging a boson star at the Galactic center}},\
  }\href@noop {} {\bibfield  {journal} {\bibinfo  {journal} {Class. Quant.
  Grav.}\ }\textbf {\bibinfo {volume} {33}},\ \bibinfo {pages} {105015}
  (\bibinfo {year} {2016})}\BibitemShut {NoStop}%
\bibitem [{\citenamefont {Grould}\ \emph {et~al.}(2017)\citenamefont {Grould},
  \citenamefont {Meliani}, \citenamefont {Vincent}, \citenamefont
  {Grandcl\'ement},\ and\ \citenamefont {Gourgoulhon}}]{Grould:2017rzz}%
  \BibitemOpen
  \bibfield  {author} {\bibinfo {author} {\bibfnamefont {M.}~\bibnamefont
  {Grould}}, \bibinfo {author} {\bibfnamefont {Z.}~\bibnamefont {Meliani}},
  \bibinfo {author} {\bibfnamefont {F.~H.}\ \bibnamefont {Vincent}}, \bibinfo
  {author} {\bibfnamefont {P.}~\bibnamefont {Grandcl\'ement}},\ and\ \bibinfo
  {author} {\bibfnamefont {E.}~\bibnamefont {Gourgoulhon}},\ }\bibfield
  {title} {\bibinfo {title} {{Comparing timelike geodesics around a Kerr black
  hole and a boson star}},\ }\href@noop {} {\bibfield  {journal} {\bibinfo
  {journal} {Class. Quant. Grav.}\ }\textbf {\bibinfo {volume} {34}},\ \bibinfo
  {pages} {215007} (\bibinfo {year} {2017})}\BibitemShut {NoStop}%
\bibitem [{\citenamefont {Olivares}\ \emph {et~al.}(2020)\citenamefont
  {Olivares}, \citenamefont {Younsi}, \citenamefont {Fromm}, \citenamefont
  {De~Laurentis}, \citenamefont {Porth}, \citenamefont {Mizuno}, \citenamefont
  {Falcke}, \citenamefont {Kramer},\ and\ \citenamefont
  {Rezzolla}}]{Olivares:2018abq}%
  \BibitemOpen
  \bibfield  {author} {\bibinfo {author} {\bibfnamefont {H.}~\bibnamefont
  {Olivares}}, \bibinfo {author} {\bibfnamefont {Z.}~\bibnamefont {Younsi}},
  \bibinfo {author} {\bibfnamefont {C.~M.}\ \bibnamefont {Fromm}}, \bibinfo
  {author} {\bibfnamefont {M.}~\bibnamefont {De~Laurentis}}, \bibinfo {author}
  {\bibfnamefont {O.}~\bibnamefont {Porth}}, \bibinfo {author} {\bibfnamefont
  {Y.}~\bibnamefont {Mizuno}}, \bibinfo {author} {\bibfnamefont
  {H.}~\bibnamefont {Falcke}}, \bibinfo {author} {\bibfnamefont
  {M.}~\bibnamefont {Kramer}},\ and\ \bibinfo {author} {\bibfnamefont
  {L.}~\bibnamefont {Rezzolla}},\ }\bibfield  {title} {\bibinfo {title} {{How
  to tell an accreting boson star from a black hole}},\ }\href@noop {}
  {\bibfield  {journal} {\bibinfo  {journal} {MNRAS}\ }\textbf {\bibinfo
  {volume} {497}},\ \bibinfo {pages} {521} (\bibinfo {year}
  {2020})}\BibitemShut {NoStop}%
\bibitem [{\citenamefont {Giddings}\ and\ \citenamefont
  {Psaltis}(2018)}]{Giddings:2016btb}%
  \BibitemOpen
  \bibfield  {author} {\bibinfo {author} {\bibfnamefont {S.~B.}\ \bibnamefont
  {Giddings}}\ and\ \bibinfo {author} {\bibfnamefont {D.}~\bibnamefont
  {Psaltis}},\ }\bibfield  {title} {\bibinfo {title} {{Event Horizon Telescope
  Observations as Probes for Quantum Structure of Astrophysical Black Holes}},\
  }\href {https://doi.org/10.1103/PhysRevD.97.084035} {\bibfield  {journal}
  {\bibinfo  {journal} {Phys. Rev. D}\ }\textbf {\bibinfo {volume} {97}},\
  \bibinfo {pages} {084035} (\bibinfo {year} {2018})},\ \Eprint
  {https://arxiv.org/abs/1606.07814} {arXiv:1606.07814 [astro-ph.HE]}
  \BibitemShut {NoStop}%
\bibitem [{\citenamefont {Bena}\ and\ \citenamefont
  {Warner}(2013)}]{Bena:2013dka}%
  \BibitemOpen
  \bibfield  {author} {\bibinfo {author} {\bibfnamefont {I.}~\bibnamefont
  {Bena}}\ and\ \bibinfo {author} {\bibfnamefont {N.~P.}\ \bibnamefont
  {Warner}},\ }\bibfield  {title} {\bibinfo {title} {{Resolving the Structure
  of Black Holes: Philosophizing with a Hammer}},\ }\href@noop {} {\bibfield
  {journal} {\bibinfo  {journal} {arXiv:1311.4538}\ } (\bibinfo {year}
  {2013})}\BibitemShut {NoStop}%
\bibitem [{\citenamefont {Bena}\ \emph {et~al.}(2016)\citenamefont {Bena},
  \citenamefont {Mayerson}, \citenamefont {Puhm},\ and\ \citenamefont
  {Vercnocke}}]{Bena:2015dpt}%
  \BibitemOpen
  \bibfield  {author} {\bibinfo {author} {\bibfnamefont {I.}~\bibnamefont
  {Bena}}, \bibinfo {author} {\bibfnamefont {D.~R.}\ \bibnamefont {Mayerson}},
  \bibinfo {author} {\bibfnamefont {A.}~\bibnamefont {Puhm}},\ and\ \bibinfo
  {author} {\bibfnamefont {B.}~\bibnamefont {Vercnocke}},\ }\bibfield  {title}
  {\bibinfo {title} {{Tunneling into Microstate Geometries: Quantum Effects
  Stop Gravitational Collapse}},\ }\href@noop {} {\bibfield  {journal}
  {\bibinfo  {journal} {JHEP}\ }\textbf {\bibinfo {volume} {07}}\bibinfo
  {number} { (2016)},\ \bibinfo {pages} {031}}\BibitemShut {NoStop}%
\bibitem [{\citenamefont {Bena}\ \emph {et~al.}(2018)\citenamefont {Bena},
  \citenamefont {Giusto}, \citenamefont {Martinec}, \citenamefont {Russo},
  \citenamefont {Shigemori}, \citenamefont {Turton},\ and\ \citenamefont
  {Warner}}]{Bena:2017xbt}%
  \BibitemOpen
\bibfield  {number} {  }\bibfield  {author} {\bibinfo {author} {\bibfnamefont
  {I.}~\bibnamefont {Bena}}, \bibinfo {author} {\bibfnamefont {S.}~\bibnamefont
  {Giusto}}, \bibinfo {author} {\bibfnamefont {E.~J.}\ \bibnamefont
  {Martinec}}, \bibinfo {author} {\bibfnamefont {R.}~\bibnamefont {Russo}},
  \bibinfo {author} {\bibfnamefont {M.}~\bibnamefont {Shigemori}}, \bibinfo
  {author} {\bibfnamefont {D.}~\bibnamefont {Turton}},\ and\ \bibinfo {author}
  {\bibfnamefont {N.~P.}\ \bibnamefont {Warner}},\ }\bibfield  {title}
  {\bibinfo {title} {{Asymptotically-flat supergravity solutions deep inside
  the black-hole regime}},\ }\href@noop {} {\bibfield  {journal} {\bibinfo
  {journal} {JHEP}\ }\textbf {\bibinfo {volume} {02}}\bibinfo  {number} {
  (2018)},\ \bibinfo {pages} {014}}\BibitemShut {NoStop}%
\bibitem [{\citenamefont {Heidmann}\ \emph {et~al.}(2020)\citenamefont
  {Heidmann}, \citenamefont {Mayerson}, \citenamefont {Walker},\ and\
  \citenamefont {Warner}}]{Heidmann:2019xrd}%
  \BibitemOpen
\bibfield  {number} {  }\bibfield  {author} {\bibinfo {author} {\bibfnamefont
  {P.}~\bibnamefont {Heidmann}}, \bibinfo {author} {\bibfnamefont {D.~R.}\
  \bibnamefont {Mayerson}}, \bibinfo {author} {\bibfnamefont {R.}~\bibnamefont
  {Walker}},\ and\ \bibinfo {author} {\bibfnamefont {N.~P.}\ \bibnamefont
  {Warner}},\ }\bibfield  {title} {\bibinfo {title} {{Holomorphic Waves of
  Black Hole Microstructure}},\ }\href@noop {} {\bibfield  {journal} {\bibinfo
  {journal} {JHEP}\ }\textbf {\bibinfo {volume} {02}}\bibinfo  {number} {
  (2020)},\ \bibinfo {pages} {192}}\BibitemShut {NoStop}%
\bibitem [{\citenamefont {Ooguri}\ and\ \citenamefont
  {Vafa}(2007)}]{Ooguri:2006in}%
  \BibitemOpen
\bibfield  {number} {  }\bibfield  {author} {\bibinfo {author} {\bibfnamefont
  {H.}~\bibnamefont {Ooguri}}\ and\ \bibinfo {author} {\bibfnamefont
  {C.}~\bibnamefont {Vafa}},\ }\bibfield  {title} {\bibinfo {title} {{On the
  Geometry of the String Landscape and the Swampland}},\ }\href
  {https://doi.org/10.1016/j.nuclphysb.2006.10.033} {\bibfield  {journal}
  {\bibinfo  {journal} {Nucl. Phys. B}\ }\textbf {\bibinfo {volume} {766}},\
  \bibinfo {pages} {21} (\bibinfo {year} {2007})},\ \Eprint
  {https://arxiv.org/abs/hep-th/0605264} {arXiv:hep-th/0605264} \BibitemShut
  {NoStop}%
\bibitem [{\citenamefont {Li}(2021)}]{Li:2021gbg}%
  \BibitemOpen
  \bibfield  {author} {\bibinfo {author} {\bibfnamefont {Y.}~\bibnamefont
  {Li}},\ }\bibfield  {title} {\bibinfo {title} {{Black holes and the
  swampland: the deep throat revelations}},\ }\href
  {https://doi.org/10.1007/JHEP06(2021)065} {\bibfield  {journal} {\bibinfo
  {journal} {JHEP}\ }\textbf {\bibinfo {volume} {06}}\bibfield  {number}
  {\bibinfo  {number} { (2021)},\ \bibinfo {pages} {065}},\ }\Eprint
  {https://arxiv.org/abs/2102.04480} {arXiv:2102.04480 [hep-th]} \BibitemShut
  {NoStop}%
\bibitem [{\citenamefont {Hertog}\ and\ \citenamefont
  {Hartle}(2020)}]{Hertog:2017vod}%
  \BibitemOpen
  \bibfield  {author} {\bibinfo {author} {\bibfnamefont {T.}~\bibnamefont
  {Hertog}}\ and\ \bibinfo {author} {\bibfnamefont {J.}~\bibnamefont
  {Hartle}},\ }\bibfield  {title} {\bibinfo {title} {Observational implications
  of fuzzball formation},\ }\href@noop {} {\bibfield  {journal} {\bibinfo
  {journal} {Gen. Rel. Grav.}\ }\textbf {\bibinfo {volume} {52,7}} (\bibinfo
  {year} {2020})}\BibitemShut {NoStop}%
\bibitem [{\citenamefont {Mayerson}(2020)}]{Mayerson:2020tpn}%
  \BibitemOpen
  \bibfield  {author} {\bibinfo {author} {\bibfnamefont {D.~R.}\ \bibnamefont
  {Mayerson}},\ }\bibfield  {title} {\bibinfo {title} {{Fuzzballs and
  Observations}},\ }\href@noop {} {\bibfield  {journal} {\bibinfo  {journal}
  {Gen. Rel. Grav.}\ }\textbf {\bibinfo {volume} {52}},\ \bibinfo {pages} {115}
  (\bibinfo {year} {2020})}\BibitemShut {NoStop}%
\bibitem [{\citenamefont {Bena}\ and\ \citenamefont
  {Mayerson}(2020{\natexlab{a}})}]{Bena:2020see}%
  \BibitemOpen
  \bibfield  {author} {\bibinfo {author} {\bibfnamefont {I.}~\bibnamefont
  {Bena}}\ and\ \bibinfo {author} {\bibfnamefont {D.~R.}\ \bibnamefont
  {Mayerson}},\ }\bibfield  {title} {\bibinfo {title} {{Multipole Ratios: A New
  Window into Black Holes}},\ }\href@noop {} {\bibfield  {journal} {\bibinfo
  {journal} {PRL}\ }\textbf {\bibinfo {volume} {125}},\ \bibinfo {pages}
  {221602} (\bibinfo {year} {2020}{\natexlab{a}})}\BibitemShut {NoStop}%
\bibitem [{\citenamefont {Bena}\ and\ \citenamefont
  {Mayerson}(2020{\natexlab{b}})}]{Bena:2020uup}%
  \BibitemOpen
  \bibfield  {author} {\bibinfo {author} {\bibfnamefont {I.}~\bibnamefont
  {Bena}}\ and\ \bibinfo {author} {\bibfnamefont {D.~R.}\ \bibnamefont
  {Mayerson}},\ }\bibfield  {title} {\bibinfo {title} {{Black Holes Lessons
  from Multipole Ratios}},\ }\href@noop {} {\bibfield  {journal} {\bibinfo
  {journal} {arXiv:2007.09152}\ } (\bibinfo {year}
  {2020}{\natexlab{b}})}\BibitemShut {NoStop}%
\bibitem [{\citenamefont {Bianchi}\ \emph {et~al.}(2020)\citenamefont
  {Bianchi}, \citenamefont {Consoli}, \citenamefont {Grillo}, \citenamefont
  {Morales}, \citenamefont {Pani},\ and\ \citenamefont
  {Raposo}}]{Bianchi:2020bxa}%
  \BibitemOpen
  \bibfield  {author} {\bibinfo {author} {\bibfnamefont {M.}~\bibnamefont
  {Bianchi}}, \bibinfo {author} {\bibfnamefont {D.}~\bibnamefont {Consoli}},
  \bibinfo {author} {\bibfnamefont {A.}~\bibnamefont {Grillo}}, \bibinfo
  {author} {\bibfnamefont {J.~F.}\ \bibnamefont {Morales}}, \bibinfo {author}
  {\bibfnamefont {P.}~\bibnamefont {Pani}},\ and\ \bibinfo {author}
  {\bibfnamefont {G.}~\bibnamefont {Raposo}},\ }\bibfield  {title} {\bibinfo
  {title} {{Distinguishing fuzzballs from black holes through their multipolar
  structure}},\ }\href@noop {} {\bibfield  {journal} {\bibinfo  {journal}
  {PRL}\ }\textbf {\bibinfo {volume} {125}},\ \bibinfo {pages} {221601}
  (\bibinfo {year} {2020})}\BibitemShut {NoStop}%
\bibitem [{\citenamefont {Bianchi}\ \emph {et~al.}(2021)\citenamefont
  {Bianchi}, \citenamefont {Consoli}, \citenamefont {Grillo}, \citenamefont
  {Morales}, \citenamefont {Pani},\ and\ \citenamefont
  {Raposo}}]{Bianchi:2020miz}%
  \BibitemOpen
  \bibfield  {author} {\bibinfo {author} {\bibfnamefont {M.}~\bibnamefont
  {Bianchi}}, \bibinfo {author} {\bibfnamefont {D.}~\bibnamefont {Consoli}},
  \bibinfo {author} {\bibfnamefont {A.}~\bibnamefont {Grillo}}, \bibinfo
  {author} {\bibfnamefont {J.~F.}\ \bibnamefont {Morales}}, \bibinfo {author}
  {\bibfnamefont {P.}~\bibnamefont {Pani}},\ and\ \bibinfo {author}
  {\bibfnamefont {G.}~\bibnamefont {Raposo}},\ }\bibfield  {title} {\bibinfo
  {title} {{The multipolar structure of fuzzballs}},\ }\href@noop {} {\bibfield
   {journal} {\bibinfo  {journal} {JHEP}\ }\textbf {\bibinfo {volume}
  {01}}\bibinfo  {number} { (2021)},\ \bibinfo {pages} {003}}\BibitemShut
  {NoStop}%
\bibitem [{\citenamefont {Bates}\ and\ \citenamefont
  {Denef}(2011)}]{Bates:2003vx}%
  \BibitemOpen
\bibfield  {number} {  }\bibfield  {author} {\bibinfo {author} {\bibfnamefont
  {B.}~\bibnamefont {Bates}}\ and\ \bibinfo {author} {\bibfnamefont
  {F.}~\bibnamefont {Denef}},\ }\bibfield  {title} {\bibinfo {title} {{Exact
  solutions for supersymmetric stationary black hole composites}},\ }\href@noop
  {} {\bibfield  {journal} {\bibinfo  {journal} {JHEP}\ }\textbf {\bibinfo
  {volume} {11}}\bibinfo  {number} { (2011)},\ \bibinfo {pages}
  {127}}\BibitemShut {NoStop}%
\bibitem [{\citenamefont {Denef}(2000)}]{Denef:2000nb}%
  \BibitemOpen
\bibfield  {number} {  }\bibfield  {author} {\bibinfo {author} {\bibfnamefont
  {F.}~\bibnamefont {Denef}},\ }\bibfield  {title} {\bibinfo {title}
  {{Supergravity flows and D-brane stability}},\ }\href
  {https://doi.org/10.1088/1126-6708/2000/08/050} {\bibfield  {journal}
  {\bibinfo  {journal} {JHEP}\ }\textbf {\bibinfo {volume} {08}}\bibfield
  {number} {\bibinfo  {number} { (2000)},\ \bibinfo {pages} {050}},\ }\Eprint
  {https://arxiv.org/abs/hep-th/0005049} {arXiv:hep-th/0005049} \BibitemShut
  {NoStop}%
\bibitem [{\citenamefont {Berglund}\ \emph {et~al.}(2006)\citenamefont
  {Berglund}, \citenamefont {Gimon},\ and\ \citenamefont
  {Levi}}]{Berglund:2005vb}%
  \BibitemOpen
  \bibfield  {author} {\bibinfo {author} {\bibfnamefont {P.}~\bibnamefont
  {Berglund}}, \bibinfo {author} {\bibfnamefont {E.~G.}\ \bibnamefont
  {Gimon}},\ and\ \bibinfo {author} {\bibfnamefont {T.~S.}\ \bibnamefont
  {Levi}},\ }\bibfield  {title} {\bibinfo {title} {{Supergravity microstates
  for BPS black holes and black rings}},\ }\href
  {https://doi.org/10.1088/1126-6708/2006/06/007} {\bibfield  {journal}
  {\bibinfo  {journal} {JHEP}\ }\textbf {\bibinfo {volume} {06}}\bibfield
  {number} {\bibinfo  {number} { (2006)},\ \bibinfo {pages} {007}},\ }\Eprint
  {https://arxiv.org/abs/hep-th/0505167} {arXiv:hep-th/0505167} \BibitemShut
  {NoStop}%
\bibitem [{\citenamefont {Gauntlett}\ and\ \citenamefont
  {Gutowski}(2005)}]{Gauntlett:2004qy}%
  \BibitemOpen
  \bibfield  {author} {\bibinfo {author} {\bibfnamefont {J.~P.}\ \bibnamefont
  {Gauntlett}}\ and\ \bibinfo {author} {\bibfnamefont {J.~B.}\ \bibnamefont
  {Gutowski}},\ }\bibfield  {title} {\bibinfo {title} {{General concentric
  black rings}},\ }\href@noop {} {\bibfield  {journal} {\bibinfo  {journal}
  {PRD}\ }\textbf {\bibinfo {volume} {71}},\ \bibinfo {pages} {045002}
  (\bibinfo {year} {2005})}\BibitemShut {NoStop}%
\bibitem [{\citenamefont {Bena}\ \emph {et~al.}(2005)\citenamefont {Bena},
  \citenamefont {Kraus},\ and\ \citenamefont {Warner}}]{Bena:2005ni}%
  \BibitemOpen
  \bibfield  {author} {\bibinfo {author} {\bibfnamefont {I.}~\bibnamefont
  {Bena}}, \bibinfo {author} {\bibfnamefont {P.}~\bibnamefont {Kraus}},\ and\
  \bibinfo {author} {\bibfnamefont {N.~P.}\ \bibnamefont {Warner}},\ }\bibfield
   {title} {\bibinfo {title} {{Black rings in Taub-NUT}},\ }\href@noop {}
  {\bibfield  {journal} {\bibinfo  {journal} {PRD}\ }\textbf {\bibinfo {volume}
  {72}},\ \bibinfo {pages} {084019} (\bibinfo {year} {2005})}\BibitemShut
  {NoStop}%
\bibitem [{\citenamefont {Denef}\ \emph {et~al.}(2012)\citenamefont {Denef},
  \citenamefont {Gaiotto}, \citenamefont {Strominger}, \citenamefont {Van~den
  Bleeken},\ and\ \citenamefont {Yin}}]{Denef:2007yt}%
  \BibitemOpen
  \bibfield  {author} {\bibinfo {author} {\bibfnamefont {F.}~\bibnamefont
  {Denef}}, \bibinfo {author} {\bibfnamefont {D.}~\bibnamefont {Gaiotto}},
  \bibinfo {author} {\bibfnamefont {A.}~\bibnamefont {Strominger}}, \bibinfo
  {author} {\bibfnamefont {D.}~\bibnamefont {Van~den Bleeken}},\ and\ \bibinfo
  {author} {\bibfnamefont {X.}~\bibnamefont {Yin}},\ }\bibfield  {title}
  {\bibinfo {title} {{Black Hole Deconstruction}},\ }\href@noop {} {\bibfield
  {journal} {\bibinfo  {journal} {JHEP}\ }\textbf {\bibinfo {volume}
  {03}}\bibinfo  {number} { (2012)},\ \bibinfo {pages} {071}}\BibitemShut
  {NoStop}%
\bibitem [{\citenamefont {Levi}\ \emph {et~al.}(2010)\citenamefont {Levi},
  \citenamefont {Raeymaekers}, \citenamefont {Van~den Bleeken}, \citenamefont
  {Van~Herck},\ and\ \citenamefont {Vercnocke}}]{Levi:2009az}%
  \BibitemOpen
\bibfield  {number} {  }\bibfield  {author} {\bibinfo {author} {\bibfnamefont
  {T.~S.}\ \bibnamefont {Levi}}, \bibinfo {author} {\bibfnamefont
  {J.}~\bibnamefont {Raeymaekers}}, \bibinfo {author} {\bibfnamefont
  {D.}~\bibnamefont {Van~den Bleeken}}, \bibinfo {author} {\bibfnamefont
  {W.}~\bibnamefont {Van~Herck}},\ and\ \bibinfo {author} {\bibfnamefont
  {B.}~\bibnamefont {Vercnocke}},\ }\bibfield  {title} {\bibinfo {title}
  {{Godel space from wrapped M2-branes}},\ }\href@noop {} {\bibfield  {journal}
  {\bibinfo  {journal} {JHEP}\ }\textbf {\bibinfo {volume} {01}}\bibinfo
  {number} { (2010)},\ \bibinfo {pages} {082}}\BibitemShut {NoStop}%
\bibitem [{\citenamefont {Raeymaekers}\ and\ \citenamefont {Van~den
  Bleeken}(2015)}]{Raeymaekers:2015sba}%
  \BibitemOpen
\bibfield  {number} {  }\bibfield  {author} {\bibinfo {author} {\bibfnamefont
  {J.}~\bibnamefont {Raeymaekers}}\ and\ \bibinfo {author} {\bibfnamefont
  {D.}~\bibnamefont {Van~den Bleeken}},\ }\bibfield  {title} {\bibinfo {title}
  {{Microstate solutions from black hole deconstruction}},\ }\href@noop {}
  {\bibfield  {journal} {\bibinfo  {journal} {JHEP}\ }\textbf {\bibinfo
  {volume} {12}}\bibinfo  {number} { (2015)},\ \bibinfo {pages}
  {095}}\BibitemShut {NoStop}%
\bibitem [{\citenamefont {Maldacena}\ \emph {et~al.}(1997)\citenamefont
  {Maldacena}, \citenamefont {Strominger},\ and\ \citenamefont
  {Witten}}]{Maldacena:1997de}%
  \BibitemOpen
\bibfield  {number} {  }\bibfield  {author} {\bibinfo {author} {\bibfnamefont
  {J.~M.}\ \bibnamefont {Maldacena}}, \bibinfo {author} {\bibfnamefont
  {A.}~\bibnamefont {Strominger}},\ and\ \bibinfo {author} {\bibfnamefont
  {E.}~\bibnamefont {Witten}},\ }\bibfield  {title} {\bibinfo {title} {{Black
  hole entropy in M theory}},\ }\href@noop {} {\bibfield  {journal} {\bibinfo
  {journal} {JHEP}\ }\textbf {\bibinfo {volume} {12}}\bibinfo  {number} {
  (1997)},\ \bibinfo {pages} {002}}\BibitemShut {NoStop}%
\bibitem [{\citenamefont {Gaiotto}\ \emph {et~al.}(2005)\citenamefont
  {Gaiotto}, \citenamefont {Strominger},\ and\ \citenamefont
  {Yin}}]{Gaiotto:2004ij}%
  \BibitemOpen
\bibfield  {number} {  }\bibfield  {author} {\bibinfo {author} {\bibfnamefont
  {D.}~\bibnamefont {Gaiotto}}, \bibinfo {author} {\bibfnamefont
  {A.}~\bibnamefont {Strominger}},\ and\ \bibinfo {author} {\bibfnamefont
  {X.}~\bibnamefont {Yin}},\ }\bibfield  {title} {\bibinfo {title}
  {{Superconformal black hole quantum mechanics}},\ }\href@noop {} {\bibfield
  {journal} {\bibinfo  {journal} {JHEP}\ }\textbf {\bibinfo {volume}
  {11}}\bibinfo  {number} { (2005)},\ \bibinfo {pages} {017}}\BibitemShut
  {NoStop}%
\bibitem [{\citenamefont {Bena}\ and\ \citenamefont
  {Warner}(2008)}]{Bena:2007kg}%
  \BibitemOpen
\bibfield  {number} {  }\bibfield  {author} {\bibinfo {author} {\bibfnamefont
  {I.}~\bibnamefont {Bena}}\ and\ \bibinfo {author} {\bibfnamefont {N.~P.}\
  \bibnamefont {Warner}},\ }\bibfield  {title} {\bibinfo {title} {{Black holes,
  black rings and their microstates}},\ }\href@noop {} {\bibfield  {journal}
  {\bibinfo  {journal} {Lect. Notes Phys.}\ }\textbf {\bibinfo {volume}
  {755}},\ \bibinfo {pages} {1} (\bibinfo {year} {2008})}\BibitemShut {NoStop}%
\bibitem [{\citenamefont {Bronzwaer}\ \emph {et~al.}(2018)\citenamefont
  {Bronzwaer}, \citenamefont {Davelaar}, \citenamefont {Younsi}, \citenamefont
  {Mo\'{s}cibrodzka}, \citenamefont {Falcke}, \citenamefont {Kramer},\ and\
  \citenamefont {Rezzolla}}]{bronzwaer2018}%
  \BibitemOpen
  \bibfield  {author} {\bibinfo {author} {\bibfnamefont {T.}~\bibnamefont
  {Bronzwaer}}, \bibinfo {author} {\bibfnamefont {J.}~\bibnamefont {Davelaar}},
  \bibinfo {author} {\bibfnamefont {Z.}~\bibnamefont {Younsi}}, \bibinfo
  {author} {\bibfnamefont {M.}~\bibnamefont {Mo\'{s}cibrodzka}}, \bibinfo
  {author} {\bibfnamefont {H.}~\bibnamefont {Falcke}}, \bibinfo {author}
  {\bibfnamefont {M.}~\bibnamefont {Kramer}},\ and\ \bibinfo {author}
  {\bibfnamefont {L.}~\bibnamefont {Rezzolla}},\ }\bibfield  {title} {\bibinfo
  {title} {{RAPTOR I: Time-dependent radiative transfer in arbitrary
  spacetimes}},\ }\href@noop {} {\bibfield  {journal} {\bibinfo  {journal}
  {A\&A}\ }\textbf {\bibinfo {volume} {613, A2}} (\bibinfo {year}
  {2018})}\BibitemShut {NoStop}%
\bibitem [{\citenamefont {Davelaar}\ \emph {et~al.}(2018)\citenamefont
  {Davelaar}, \citenamefont {Bronzwaer}, \citenamefont {Kok}, \citenamefont
  {Younsi}, \citenamefont {Mo\'{s}cibrodzka},\ and\ \citenamefont
  {Falcke}}]{davelaar2018b}%
  \BibitemOpen
  \bibfield  {author} {\bibinfo {author} {\bibfnamefont {J.}~\bibnamefont
  {Davelaar}}, \bibinfo {author} {\bibfnamefont {T.}~\bibnamefont {Bronzwaer}},
  \bibinfo {author} {\bibfnamefont {D.}~\bibnamefont {Kok}}, \bibinfo {author}
  {\bibfnamefont {Z.}~\bibnamefont {Younsi}}, \bibinfo {author} {\bibfnamefont
  {M.}~\bibnamefont {Mo\'{s}cibrodzka}},\ and\ \bibinfo {author} {\bibfnamefont
  {H.}~\bibnamefont {Falcke}},\ }\bibfield  {title} {\bibinfo {title}
  {Observing supermassive black holes in virtual reality},\ }\href@noop {}
  {\bibfield  {journal} {\bibinfo  {journal} {Comput. Astrophys. Cosm.}\
  }\textbf {\bibinfo {volume} {5, 1}} (\bibinfo {year} {2018})}\BibitemShut
  {NoStop}%
\bibitem [{\citenamefont {Bronzwaer}\ \emph {et~al.}(2020)\citenamefont
  {Bronzwaer}, \citenamefont {Younsi}, \citenamefont {Davelaar},\ and\
  \citenamefont {Falcke}}]{bronzwaer2020}%
  \BibitemOpen
  \bibfield  {author} {\bibinfo {author} {\bibfnamefont {T.}~\bibnamefont
  {Bronzwaer}}, \bibinfo {author} {\bibfnamefont {Z.}~\bibnamefont {Younsi}},
  \bibinfo {author} {\bibfnamefont {J.}~\bibnamefont {Davelaar}},\ and\
  \bibinfo {author} {\bibfnamefont {H.}~\bibnamefont {Falcke}},\ }\bibfield
  {title} {\bibinfo {title} {{RAPTOR II. Polarized radiative transfer in curved
  spacetime}},\ }\href@noop {} {\bibfield  {journal} {\bibinfo  {journal}
  {A\&A}\ }\textbf {\bibinfo {volume} {641, A126}} (\bibinfo {year}
  {2020})}\BibitemShut {NoStop}%
\bibitem [{\citenamefont {Bacchini}\ \emph {et~al.}(2018)\citenamefont
  {Bacchini}, \citenamefont {Ripperda}, \citenamefont {Chen},\ and\
  \citenamefont {Sironi}}]{bacchini2018a}%
  \BibitemOpen
  \bibfield  {author} {\bibinfo {author} {\bibfnamefont {F.}~\bibnamefont
  {Bacchini}}, \bibinfo {author} {\bibfnamefont {B.}~\bibnamefont {Ripperda}},
  \bibinfo {author} {\bibfnamefont {A.}~\bibnamefont {Chen}},\ and\ \bibinfo
  {author} {\bibfnamefont {L.}~\bibnamefont {Sironi}},\ }\bibfield  {title}
  {\bibinfo {title} {{Generalized, energy-conserving numerical simulations of
  particles in general relativity. {I}. Time-like and null geodesics}},\
  }\href@noop {} {\bibfield  {journal} {\bibinfo  {journal} {ApJS}\ }\textbf
  {\bibinfo {volume} {237, 6}} (\bibinfo {year} {2018})}\BibitemShut {NoStop}%
\bibitem [{\citenamefont {Bohn}\ \emph {et~al.}(2015)\citenamefont {Bohn},
  \citenamefont {Throwe}, \citenamefont {H\'ebert}, \citenamefont {Henriksson},
  \citenamefont {Bunandar}, \citenamefont {Scheel},\ and\ \citenamefont
  {Taylor}}]{Bohn:2014xxa}%
  \BibitemOpen
  \bibfield  {author} {\bibinfo {author} {\bibfnamefont {A.}~\bibnamefont
  {Bohn}}, \bibinfo {author} {\bibfnamefont {W.}~\bibnamefont {Throwe}},
  \bibinfo {author} {\bibfnamefont {F.}~\bibnamefont {H\'ebert}}, \bibinfo
  {author} {\bibfnamefont {K.}~\bibnamefont {Henriksson}}, \bibinfo {author}
  {\bibfnamefont {D.}~\bibnamefont {Bunandar}}, \bibinfo {author}
  {\bibfnamefont {M.~A.}\ \bibnamefont {Scheel}},\ and\ \bibinfo {author}
  {\bibfnamefont {N.~W.}\ \bibnamefont {Taylor}},\ }\bibfield  {title}
  {\bibinfo {title} {{What does a binary black hole merger look like?}},\
  }\href {https://doi.org/10.1088/0264-9381/32/6/065002} {\bibfield  {journal}
  {\bibinfo  {journal} {Class. Quant. Grav.}\ }\textbf {\bibinfo {volume}
  {32}},\ \bibinfo {pages} {065002} (\bibinfo {year} {2015})},\ \Eprint
  {https://arxiv.org/abs/1410.7775} {arXiv:1410.7775 [gr-qc]} \BibitemShut
  {NoStop}%
\bibitem [{\citenamefont {Mayerson}\ \emph {et~al.}(2021)\citenamefont
  {Mayerson}, \citenamefont {Bacchini}, \citenamefont {Ripperda}, \citenamefont
  {Davelaar}, \citenamefont {Olivares}, \citenamefont {Vercnocke},\ and\
  \citenamefont {Hertog}}]{ourfollowup}%
  \BibitemOpen
  \bibfield  {author} {\bibinfo {author} {\bibfnamefont {D.~R.}\ \bibnamefont
  {Mayerson}}, \bibinfo {author} {\bibfnamefont {F.}~\bibnamefont {Bacchini}},
  \bibinfo {author} {\bibfnamefont {B.}~\bibnamefont {Ripperda}}, \bibinfo
  {author} {\bibfnamefont {J.}~\bibnamefont {Davelaar}}, \bibinfo {author}
  {\bibfnamefont {H.}~\bibnamefont {Olivares}}, \bibinfo {author}
  {\bibfnamefont {B.}~\bibnamefont {Vercnocke}},\ and\ \bibinfo {author}
  {\bibfnamefont {T.}~\bibnamefont {Hertog}},\ }\href@noop {} {\bibinfo {title}
  {in preparation}} (\bibinfo {year} {2021})\BibitemShut {NoStop}%
\bibitem [{\citenamefont {Tyukov}\ \emph {et~al.}(2018)\citenamefont {Tyukov},
  \citenamefont {Walker},\ and\ \citenamefont {Warner}}]{Tyukov:2017uig}%
  \BibitemOpen
  \bibfield  {author} {\bibinfo {author} {\bibfnamefont {A.}~\bibnamefont
  {Tyukov}}, \bibinfo {author} {\bibfnamefont {R.}~\bibnamefont {Walker}},\
  and\ \bibinfo {author} {\bibfnamefont {N.~P.}\ \bibnamefont {Warner}},\
  }\bibfield  {title} {\bibinfo {title} {{Tidal Stresses and Energy Gaps in
  Microstate Geometries}},\ }\href@noop {} {\bibfield  {journal} {\bibinfo
  {journal} {JHEP}\ }\textbf {\bibinfo {volume} {02}}\bibinfo  {number} {
  (2018)},\ \bibinfo {pages} {122}}\BibitemShut {NoStop}%
\bibitem [{\citenamefont {Bena}\ \emph {et~al.}(2019)\citenamefont {Bena},
  \citenamefont {Martinec}, \citenamefont {Walker},\ and\ \citenamefont
  {Warner}}]{Bena:2018mpb}%
  \BibitemOpen
\bibfield  {number} {  }\bibfield  {author} {\bibinfo {author} {\bibfnamefont
  {I.}~\bibnamefont {Bena}}, \bibinfo {author} {\bibfnamefont {E.~J.}\
  \bibnamefont {Martinec}}, \bibinfo {author} {\bibfnamefont {R.}~\bibnamefont
  {Walker}},\ and\ \bibinfo {author} {\bibfnamefont {N.~P.}\ \bibnamefont
  {Warner}},\ }\bibfield  {title} {\bibinfo {title} {{Early Scrambling and
  Capped BTZ Geometries}},\ }\href@noop {} {\bibfield  {journal} {\bibinfo
  {journal} {JHEP}\ }\textbf {\bibinfo {volume} {04}}\bibinfo  {number} {
  (2019)},\ \bibinfo {pages} {126}}\BibitemShut {NoStop}%
\bibitem [{\citenamefont {Bena}\ \emph {et~al.}(2020)\citenamefont {Bena},
  \citenamefont {Houppe},\ and\ \citenamefont {Warner}}]{Bena:2020iyw}%
  \BibitemOpen
\bibfield  {number} {  }\bibfield  {author} {\bibinfo {author} {\bibfnamefont
  {I.}~\bibnamefont {Bena}}, \bibinfo {author} {\bibfnamefont {A.}~\bibnamefont
  {Houppe}},\ and\ \bibinfo {author} {\bibfnamefont {N.~P.}\ \bibnamefont
  {Warner}},\ }\bibfield  {title} {\bibinfo {title} {{Delaying the Inevitable:
  Tidal Disruption in Microstate Geometries}},\ }\href@noop {} {\bibfield
  {journal} {\bibinfo  {journal} {arXiv:2006.13939}\ } (\bibinfo {year}
  {2020})}\BibitemShut {NoStop}%
\bibitem [{\citenamefont {Martinec}\ and\ \citenamefont
  {Warner}(2020)}]{Martinec:2020cml}%
  \BibitemOpen
  \bibfield  {author} {\bibinfo {author} {\bibfnamefont {E.~J.}\ \bibnamefont
  {Martinec}}\ and\ \bibinfo {author} {\bibfnamefont {N.~P.}\ \bibnamefont
  {Warner}},\ }\bibfield  {title} {\bibinfo {title} {{The Harder They Fall, the
  Bigger They Become: Tidal Trapping of Strings by Microstate Geometries}},\
  }\href@noop {} {\bibfield  {journal} {\bibinfo  {journal} {arXiv:2009.07847}\
  } (\bibinfo {year} {2020})}\BibitemShut {NoStop}%
\bibitem [{\citenamefont {Bena}\ \emph {et~al.}(2006)\citenamefont {Bena},
  \citenamefont {Wang},\ and\ \citenamefont {Warner}}]{Bena:2006kb}%
  \BibitemOpen
  \bibfield  {author} {\bibinfo {author} {\bibfnamefont {I.}~\bibnamefont
  {Bena}}, \bibinfo {author} {\bibfnamefont {C.-W.}\ \bibnamefont {Wang}},\
  and\ \bibinfo {author} {\bibfnamefont {N.~P.}\ \bibnamefont {Warner}},\
  }\bibfield  {title} {\bibinfo {title} {{Mergers and typical black hole
  microstates}},\ }\href@noop {} {\bibfield  {journal} {\bibinfo  {journal}
  {JHEP}\ }\textbf {\bibinfo {volume} {11}}\bibinfo  {number} { (2006)},\
  \bibinfo {pages} {042}}\BibitemShut {NoStop}%
\bibitem [{\citenamefont {Bena}\ \emph {et~al.}(2008)\citenamefont {Bena},
  \citenamefont {Wang},\ and\ \citenamefont {Warner}}]{Bena:2007qc}%
  \BibitemOpen
\bibfield  {number} {  }\bibfield  {author} {\bibinfo {author} {\bibfnamefont
  {I.}~\bibnamefont {Bena}}, \bibinfo {author} {\bibfnamefont {C.-W.}\
  \bibnamefont {Wang}},\ and\ \bibinfo {author} {\bibfnamefont {N.~P.}\
  \bibnamefont {Warner}},\ }\bibfield  {title} {\bibinfo {title} {{Plumbing the
  Abyss: Black ring microstates}},\ }\href@noop {} {\bibfield  {journal}
  {\bibinfo  {journal} {JHEP}\ }\textbf {\bibinfo {volume} {07}}\bibinfo
  {number} { (2008)},\ \bibinfo {pages} {019}}\BibitemShut {NoStop}%
\bibitem [{\citenamefont {Bianchi}\ \emph {et~al.}(2018)\citenamefont
  {Bianchi}, \citenamefont {Consoli},\ and\ \citenamefont
  {Morales}}]{Bianchi:2017sds}%
  \BibitemOpen
\bibfield  {number} {  }\bibfield  {author} {\bibinfo {author} {\bibfnamefont
  {M.}~\bibnamefont {Bianchi}}, \bibinfo {author} {\bibfnamefont
  {D.}~\bibnamefont {Consoli}},\ and\ \bibinfo {author} {\bibfnamefont {J.~F.}\
  \bibnamefont {Morales}},\ }\bibfield  {title} {\bibinfo {title} {{Probing
  Fuzzballs with Particles, Waves and Strings}},\ }\href@noop {} {\bibfield
  {journal} {\bibinfo  {journal} {JHEP}\ }\textbf {\bibinfo {volume}
  {06}}\bibinfo  {number} { (2018)},\ \bibinfo {pages} {157}}\BibitemShut
  {NoStop}%
\bibitem [{\citenamefont {Bianchi}\ \emph {et~al.}(2019)\citenamefont
  {Bianchi}, \citenamefont {Consoli}, \citenamefont {Grillo},\ and\
  \citenamefont {Morales}}]{Bianchi:2018kzy}%
  \BibitemOpen
\bibfield  {number} {  }\bibfield  {author} {\bibinfo {author} {\bibfnamefont
  {M.}~\bibnamefont {Bianchi}}, \bibinfo {author} {\bibfnamefont
  {D.}~\bibnamefont {Consoli}}, \bibinfo {author} {\bibfnamefont
  {A.}~\bibnamefont {Grillo}},\ and\ \bibinfo {author} {\bibfnamefont {J.~F.}\
  \bibnamefont {Morales}},\ }\bibfield  {title} {\bibinfo {title} {{The dark
  side of fuzzball geometries}},\ }\href@noop {} {\bibfield  {journal}
  {\bibinfo  {journal} {JHEP}\ }\textbf {\bibinfo {volume} {05}}\bibinfo
  {number} { (2019)},\ \bibinfo {pages} {126}}\BibitemShut {NoStop}%
\bibitem [{\citenamefont {Eperon}\ \emph {et~al.}(2016)\citenamefont {Eperon},
  \citenamefont {Reall},\ and\ \citenamefont {Santos}}]{Eperon:2016cdd}%
  \BibitemOpen
\bibfield  {number} {  }\bibfield  {author} {\bibinfo {author} {\bibfnamefont
  {F.~C.}\ \bibnamefont {Eperon}}, \bibinfo {author} {\bibfnamefont {H.~S.}\
  \bibnamefont {Reall}},\ and\ \bibinfo {author} {\bibfnamefont {J.~E.}\
  \bibnamefont {Santos}},\ }\bibfield  {title} {\bibinfo {title} {{Instability
  of supersymmetric microstate geometries}},\ }\href@noop {} {\bibfield
  {journal} {\bibinfo  {journal} {JHEP}\ }\textbf {\bibinfo {volume}
  {10}}\bibinfo  {number} { (2016)},\ \bibinfo {pages} {031}}\BibitemShut
  {NoStop}%
\bibitem [{\citenamefont {Keir}(2019)}]{Keir:2016azt}%
  \BibitemOpen
\bibfield  {number} {  }\bibfield  {author} {\bibinfo {author} {\bibfnamefont
  {J.}~\bibnamefont {Keir}},\ }\bibfield  {title} {\bibinfo {title} {{Wave
  propagation on microstate geometries}},\ }\href@noop {} {\bibfield  {journal}
  {\bibinfo  {journal} {Annales Henri Poincare}\ }\textbf {\bibinfo {volume}
  {21}},\ \bibinfo {pages} {705} (\bibinfo {year} {2019})}\BibitemShut
  {NoStop}%
\bibitem [{\citenamefont {Marolf}\ \emph {et~al.}(2017)\citenamefont {Marolf},
  \citenamefont {Michel},\ and\ \citenamefont {Puhm}}]{Marolf:2016nwu}%
  \BibitemOpen
  \bibfield  {author} {\bibinfo {author} {\bibfnamefont {D.}~\bibnamefont
  {Marolf}}, \bibinfo {author} {\bibfnamefont {B.}~\bibnamefont {Michel}},\
  and\ \bibinfo {author} {\bibfnamefont {A.}~\bibnamefont {Puhm}},\ }\bibfield
  {title} {\bibinfo {title} {{A rough end for smooth microstate geometries}},\
  }\href@noop {} {\bibfield  {journal} {\bibinfo  {journal} {JHEP}\ }\textbf
  {\bibinfo {volume} {05}}\bibinfo  {number} { (2017)},\ \bibinfo {pages}
  {021}}\BibitemShut {NoStop}%
\bibitem [{\citenamefont {Eperon}(2017)}]{Eperon:2017bwq}%
  \BibitemOpen
\bibfield  {number} {  }\bibfield  {author} {\bibinfo {author} {\bibfnamefont
  {F.~C.}\ \bibnamefont {Eperon}},\ }\bibfield  {title} {\bibinfo {title}
  {{Geodesics in supersymmetric microstate geometries}},\ }\href@noop {}
  {\bibfield  {journal} {\bibinfo  {journal} {Class. Quant. Grav.}\ }\textbf
  {\bibinfo {volume} {34}},\ \bibinfo {pages} {165003} (\bibinfo {year}
  {2017})}\BibitemShut {NoStop}%
\bibitem [{\citenamefont {Cunha}\ \emph {et~al.}(2015)\citenamefont {Cunha},
  \citenamefont {Herdeiro}, \citenamefont {Radu},\ and\ \citenamefont
  {Runarsson}}]{Cunha:2015yba}%
  \BibitemOpen
  \bibfield  {author} {\bibinfo {author} {\bibfnamefont {P.~V.~P.}\
  \bibnamefont {Cunha}}, \bibinfo {author} {\bibfnamefont {C.~A.~R.}\
  \bibnamefont {Herdeiro}}, \bibinfo {author} {\bibfnamefont {E.}~\bibnamefont
  {Radu}},\ and\ \bibinfo {author} {\bibfnamefont {H.~F.}\ \bibnamefont
  {Runarsson}},\ }\bibfield  {title} {\bibinfo {title} {{Shadows of Kerr black
  holes with scalar hair}},\ }\href
  {https://doi.org/10.1103/PhysRevLett.115.211102} {\bibfield  {journal}
  {\bibinfo  {journal} {Phys. Rev. Lett.}\ }\textbf {\bibinfo {volume} {115}},\
  \bibinfo {pages} {211102} (\bibinfo {year} {2015})},\ \Eprint
  {https://arxiv.org/abs/1509.00021} {arXiv:1509.00021 [gr-qc]} \BibitemShut
  {NoStop}%
\bibitem [{\citenamefont {Cunha}\ \emph {et~al.}(2016)\citenamefont {Cunha},
  \citenamefont {Grover}, \citenamefont {Herdeiro}, \citenamefont {Radu},
  \citenamefont {Runarsson},\ and\ \citenamefont {Wittig}}]{Cunha:2016bjh}%
  \BibitemOpen
  \bibfield  {author} {\bibinfo {author} {\bibfnamefont {P.~V.~P.}\
  \bibnamefont {Cunha}}, \bibinfo {author} {\bibfnamefont {J.}~\bibnamefont
  {Grover}}, \bibinfo {author} {\bibfnamefont {C.}~\bibnamefont {Herdeiro}},
  \bibinfo {author} {\bibfnamefont {E.}~\bibnamefont {Radu}}, \bibinfo {author}
  {\bibfnamefont {H.}~\bibnamefont {Runarsson}},\ and\ \bibinfo {author}
  {\bibfnamefont {A.}~\bibnamefont {Wittig}},\ }\bibfield  {title} {\bibinfo
  {title} {{Chaotic lensing around boson stars and Kerr black holes with scalar
  hair}},\ }\href {https://doi.org/10.1103/PhysRevD.94.104023} {\bibfield
  {journal} {\bibinfo  {journal} {Phys. Rev. D}\ }\textbf {\bibinfo {volume}
  {94}},\ \bibinfo {pages} {104023} (\bibinfo {year} {2016})},\ \Eprint
  {https://arxiv.org/abs/1609.01340} {arXiv:1609.01340 [gr-qc]} \BibitemShut
  {NoStop}%
\bibitem [{\citenamefont {Cardoso}\ \emph {et~al.}(2016)\citenamefont
  {Cardoso}, \citenamefont {Macedo}, \citenamefont {Pani},\ and\ \citenamefont
  {Ferrari}}]{Cardoso:2016olt}%
  \BibitemOpen
  \bibfield  {author} {\bibinfo {author} {\bibfnamefont {V.}~\bibnamefont
  {Cardoso}}, \bibinfo {author} {\bibfnamefont {C.~F.}\ \bibnamefont {Macedo}},
  \bibinfo {author} {\bibfnamefont {P.}~\bibnamefont {Pani}},\ and\ \bibinfo
  {author} {\bibfnamefont {V.}~\bibnamefont {Ferrari}},\ }\bibfield  {title}
  {\bibinfo {title} {Black holes and gravitational waves in models of
  minicharged dark matter},\ }\href@noop {} {\bibfield  {journal} {\bibinfo
  {journal} {Journal of Cosmology and Astroparticle Physics}\ }\textbf
  {\bibinfo {volume} {2016}}\bibinfo  {number} { (05)},\ \bibinfo {pages}
  {054}}\BibitemShut {NoStop}%
\bibitem [{\citenamefont {Bozzola}\ and\ \citenamefont
  {Paschalidis}(2021)}]{Bozzola:2020mjx}%
  \BibitemOpen
\bibfield  {number} {  }\bibfield  {author} {\bibinfo {author} {\bibfnamefont
  {G.}~\bibnamefont {Bozzola}}\ and\ \bibinfo {author} {\bibfnamefont
  {V.}~\bibnamefont {Paschalidis}},\ }\bibfield  {title} {\bibinfo {title}
  {{General relativistic simulations of the quasi-circular inspiral and merger
  of charged black holes: GW150914 and fundamental physics implications}},\
  }\href@noop {} {\bibfield  {journal} {\bibinfo  {journal} {PRL}\ }\textbf
  {\bibinfo {volume} {126}},\ \bibinfo {pages} {041103} (\bibinfo {year}
  {2021})}\BibitemShut {NoStop}%
\bibitem [{\citenamefont {Ripperda}\ \emph {et~al.}(2021)\citenamefont
  {Ripperda}, \citenamefont {Davelaar}, \citenamefont {Olivares}, \citenamefont
  {Mayerson}, \citenamefont {Bacchini}, \citenamefont {Vercnocke},\ and\
  \citenamefont {Hertog}}]{ourfollowupKN}%
  \BibitemOpen
  \bibfield  {author} {\bibinfo {author} {\bibfnamefont {B.}~\bibnamefont
  {Ripperda}}, \bibinfo {author} {\bibfnamefont {J.}~\bibnamefont {Davelaar}},
  \bibinfo {author} {\bibfnamefont {H.}~\bibnamefont {Olivares}}, \bibinfo
  {author} {\bibfnamefont {D.~R.}\ \bibnamefont {Mayerson}}, \bibinfo {author}
  {\bibfnamefont {F.}~\bibnamefont {Bacchini}}, \bibinfo {author}
  {\bibfnamefont {B.}~\bibnamefont {Vercnocke}},\ and\ \bibinfo {author}
  {\bibfnamefont {T.}~\bibnamefont {Hertog}},\ }\href@noop {} {\bibinfo {title}
  {Can we detect signatures of dark matter and hidden dimensions in black-hole
  shadows?}} (\bibinfo {year} {2021}),\ \bibinfo {note} {in
  preparation}\BibitemShut {NoStop}%
\bibitem [{\citenamefont {Bueno}\ \emph {et~al.}(2018)\citenamefont {Bueno},
  \citenamefont {Cano}, \citenamefont {Goelen}, \citenamefont {Hertog},\ and\
  \citenamefont {Vercnocke}}]{Bueno:2017hyj}%
  \BibitemOpen
  \bibfield  {author} {\bibinfo {author} {\bibfnamefont {P.}~\bibnamefont
  {Bueno}}, \bibinfo {author} {\bibfnamefont {P.~A.}\ \bibnamefont {Cano}},
  \bibinfo {author} {\bibfnamefont {F.}~\bibnamefont {Goelen}}, \bibinfo
  {author} {\bibfnamefont {T.}~\bibnamefont {Hertog}},\ and\ \bibinfo {author}
  {\bibfnamefont {B.}~\bibnamefont {Vercnocke}},\ }\bibfield  {title} {\bibinfo
  {title} {{Echoes of Kerr-like wormholes}},\ }\href
  {https://doi.org/10.1103/PhysRevD.97.024040} {\bibfield  {journal} {\bibinfo
  {journal} {Phys. Rev. D}\ }\textbf {\bibinfo {volume} {97}},\ \bibinfo
  {pages} {024040} (\bibinfo {year} {2018})},\ \Eprint
  {https://arxiv.org/abs/1711.00391} {arXiv:1711.00391 [gr-qc]} \BibitemShut
  {NoStop}%
\bibitem [{\citenamefont {{EHT~Collaboration}}(2019{\natexlab{b}})}]{EHT2019e}%
  \BibitemOpen
  \bibfield  {author} {\bibinfo {author} {\bibnamefont {{EHT~Collaboration}}},\
  }\bibfield  {title} {\bibinfo {title} {{First M87 Event Horizon Telescope
  Results. V. Physical Origin of the Asymmetric Ring}},\ }\href@noop {}
  {\bibfield  {journal} {\bibinfo  {journal} {ApJL}\ }\textbf {\bibinfo
  {volume} {875, L5}} (\bibinfo {year} {2019}{\natexlab{b}})}\BibitemShut
  {NoStop}%
\bibitem [{\citenamefont {Mizuno}\ \emph {et~al.}(2018)\citenamefont {Mizuno},
  \citenamefont {Younsi}, \citenamefont {Fromm}, \citenamefont {Porth},
  \citenamefont {De~Laurentis}, \citenamefont {Olivares}, \citenamefont
  {Falcke}, \citenamefont {Kramer},\ and\ \citenamefont
  {Rezzolla}}]{mizuno2018}%
  \BibitemOpen
  \bibfield  {author} {\bibinfo {author} {\bibfnamefont {Y.}~\bibnamefont
  {Mizuno}}, \bibinfo {author} {\bibfnamefont {Z.}~\bibnamefont {Younsi}},
  \bibinfo {author} {\bibfnamefont {C.}~\bibnamefont {Fromm}}, \bibinfo
  {author} {\bibfnamefont {O.}~\bibnamefont {Porth}}, \bibinfo {author}
  {\bibfnamefont {M.}~\bibnamefont {De~Laurentis}}, \bibinfo {author}
  {\bibfnamefont {H.}~\bibnamefont {Olivares}}, \bibinfo {author}
  {\bibfnamefont {H.}~\bibnamefont {Falcke}}, \bibinfo {author} {\bibfnamefont
  {M.}~\bibnamefont {Kramer}},\ and\ \bibinfo {author} {\bibfnamefont
  {L.}~\bibnamefont {Rezzolla}},\ }\bibfield  {title} {\bibinfo {title} {The
  current ability to test theories of gravity with black hole shadows},\
  }\href@noop {} {\bibfield  {journal} {\bibinfo  {journal} {Nature Astron.}\
  }\textbf {\bibinfo {volume} {2, 585}} (\bibinfo {year} {2018})}\BibitemShut
  {NoStop}%
\bibitem [{\citenamefont {Arkani-Hamed}\ \emph {et~al.}(2007)\citenamefont
  {Arkani-Hamed}, \citenamefont {Motl}, \citenamefont {Nicolis},\ and\
  \citenamefont {Vafa}}]{Arkani-Hamed:2006emk}%
  \BibitemOpen
  \bibfield  {author} {\bibinfo {author} {\bibfnamefont {N.}~\bibnamefont
  {Arkani-Hamed}}, \bibinfo {author} {\bibfnamefont {L.}~\bibnamefont {Motl}},
  \bibinfo {author} {\bibfnamefont {A.}~\bibnamefont {Nicolis}},\ and\ \bibinfo
  {author} {\bibfnamefont {C.}~\bibnamefont {Vafa}},\ }\bibfield  {title}
  {\bibinfo {title} {{The String landscape, black holes and gravity as the
  weakest force}},\ }\href {https://doi.org/10.1088/1126-6708/2007/06/060}
  {\bibfield  {journal} {\bibinfo  {journal} {JHEP}\ }\textbf {\bibinfo
  {volume} {06}}\bibfield  {number} {\bibinfo  {number} { (2007)},\ \bibinfo
  {pages} {060}},\ }\Eprint {https://arxiv.org/abs/hep-th/0601001}
  {arXiv:hep-th/0601001} \BibitemShut {NoStop}%
\bibitem [{\citenamefont {Loges}\ \emph {et~al.}(2020)\citenamefont {Loges},
  \citenamefont {Noumi},\ and\ \citenamefont {Shiu}}]{Loges:2019jzs}%
  \BibitemOpen
  \bibfield  {author} {\bibinfo {author} {\bibfnamefont {G.~J.}\ \bibnamefont
  {Loges}}, \bibinfo {author} {\bibfnamefont {T.}~\bibnamefont {Noumi}},\ and\
  \bibinfo {author} {\bibfnamefont {G.}~\bibnamefont {Shiu}},\ }\bibfield
  {title} {\bibinfo {title} {{Thermodynamics of 4D Dilatonic Black Holes and
  the Weak Gravity Conjecture}},\ }\href
  {https://doi.org/10.1103/PhysRevD.102.046010} {\bibfield  {journal} {\bibinfo
   {journal} {Phys. Rev. D}\ }\textbf {\bibinfo {volume} {102}},\ \bibinfo
  {pages} {046010} (\bibinfo {year} {2020})},\ \Eprint
  {https://arxiv.org/abs/1909.01352} {arXiv:1909.01352 [hep-th]} \BibitemShut
  {NoStop}%
\bibitem [{\citenamefont {Dimitrov}\ \emph {et~al.}(2020)\citenamefont
  {Dimitrov}, \citenamefont {Lemmens}, \citenamefont {Mayerson}, \citenamefont
  {Min},\ and\ \citenamefont {Vercnocke}}]{Dimitrov:2020txx}%
  \BibitemOpen
  \bibfield  {author} {\bibinfo {author} {\bibfnamefont {V.}~\bibnamefont
  {Dimitrov}}, \bibinfo {author} {\bibfnamefont {T.}~\bibnamefont {Lemmens}},
  \bibinfo {author} {\bibfnamefont {D.}~\bibnamefont {Mayerson}}, \bibinfo
  {author} {\bibfnamefont {V.}~\bibnamefont {Min}},\ and\ \bibinfo {author}
  {\bibfnamefont {B.}~\bibnamefont {Vercnocke}},\ }\bibfield  {title} {\bibinfo
  {title} {{Gravitational Waves, Holography, and Black Hole Microstates}},\
  }\href@noop {} {\bibfield  {journal} {\bibinfo  {journal} {arXiv:2007.01879}\
  } (\bibinfo {year} {2020})}\BibitemShut {NoStop}%
\end{thebibliography}
%

\end{document}